\documentclass[aps,pra,twocolumn,floats,amsmath,amssymb,superscriptaddress]{revtex4-1}
\usepackage[utf8]{inputenc}
\usepackage{graphicx}
\usepackage{epsfig}
\usepackage{amsfonts}
\usepackage{amsmath}
\usepackage{natbib}
\usepackage{multirow}
\usepackage{bm}
\usepackage{epstopdf}
\usepackage{ulem}
\usepackage{color}
\usepackage{hyperref}
\newcommand{\ie}{{\it i.e.}, }

\newcommand{\exciting}{{\usefont{T1}{lmtt}{b}{n}exciting}}
\begin{document}
\title{Hybrid excitations at the interface between a MoS$_2$ monolayer and organic molecules: a first-principles study}
\author{Ignacio Gonzalez Oliva}
\affiliation{Institut f\"{u}r Physik and IRIS Adlershof, Humboldt-Universit\"{a}t zu Berlin, 12489 Berlin, Germany}

\author{Fabio Caruso}
\affiliation{Institut f\"{u}r Physik and IRIS Adlershof, Humboldt-Universit\"{a}t zu Berlin, 12489 Berlin, Germany}
\affiliation{Institut f\"{u}r Theoretische Physik und Astrophysik, Christian-Albrechts-Universit\"{a}t zu Kiel, Kiel, Germany}

\author{Pasquale Pavone}
\affiliation{Institut f\"{u}r Physik and IRIS Adlershof, Humboldt-Universit\"{a}t zu Berlin, 12489 Berlin, Germany}
\affiliation{European Theoretical Spectroscopic Facility (ETSF)}

\author{Claudia Draxl}
\affiliation{Institut f\"{u}r Physik and IRIS Adlershof, Humboldt-Universit\"{a}t zu Berlin, 12489 Berlin, Germany}
\affiliation{European Theoretical Spectroscopic Facility (ETSF)}

\date{\today}
\begin{abstract}
We present a first-principles investigation of the electronic and optical properties of hybrid organic-inorganic interfaces consisting of MoS$_2$ monolayer and the $\pi$-conjugate molecules pyrene and pyridine. For both hybrid systems, the quasi-particle band structure obtained from the $G_0W_0$ approximation shows -- in contrast to density-functional theory -- level alignment of type~II, owing to the mutual dynamical screening of the interface constituents. {\it Ab initio} calculations of the absorption spectrum based on the Bethe-Salpeter equation reveal besides intra-layer excitons on the MoS$_2$ side, hybrid as well as charge-transfer excitons at the interface. These findings indicate that hybrid systems consisting of semiconducting transition-metal dichalcogenides and organic $\pi$-conjugate molecules can host a rich variety of optical excitations and thus provide a promising venue to explore many-body interactions and exciton physics in low dimensionality. 
\end{abstract}
\maketitle
\section{Introduction}

Hybrid materials composed out of organic and two-dimensional (2D) inorganic semiconductors are receiving increasing attention due to their potential for applications in opto-electronic devices, such as photodetectors, photovoltaic absorbers, and nanoscale transistors \citep{Klots2014, Lembke2015, Manzeli2017, Kwon2019}. 

The interplay of orbital hybridization, charge transfer, and dipole moments at the interface between organic molecules and inorganic substrates \citep{Romaner2008, Garcia-Lastra2009, Puschnig2012, Schlesinger2015, Otero2017} can profoundly influence the material's response to external perturbations, providing a unique degree of freedom to tailor light-matter interaction.

Semiconducting few-layer transition-metal dichalcogenides (TMDCs), such as a MoS$_2$ monolayer, exhibit a direct band gap in the visible range and high electron mobility \citep{Molina-Sanchez2013, Molina2015, Qiu2013,  Qiu2016, Hagara2020, Caruso2021}, and they thus represent promising candidates for the exploration of novel classes of hybrid materials \citep{Liu2017, Huang2018, Park2021}. Organic molecules, assembled on the surface of TMDCs through dipole or van der Waals (vdW) interactions, can further lead to unique physical properties and processes that are absent in three-dimensional semiconductors \citep{Jing2014, Wang2020, Mutz2020, Canton-Vitoria2020, Yousofnejad2020}. For example, recent studies revealed an enhancement in the photo-response of MoS$_2$ in presence of physisorbed phthalocyanin molecules \citep{Mutz2020}. Pyrene molecules, in turn, have been used as coating to protect TMDCs from photooxidation and environmental aging \citep{Canton-Vitoria2020}. Moreover, it has been demonstrated that MoS$_2$ may act as a decoupling layer on the electronic and vibrational properties of tetracyanoquinodimethane (TCNQ) \citep{Yousofnejad2020}. 

First-principles calculations based on density-functional theory (DFT) and many-body perturbation theory --in terms of the $GW$ approximation \cite{Hybertsen1986} and the Bethe-Salpeter equation (BSE) \citep{Rohlfing2000}-- provide a versatile and accurate framework to compute the electronic structure of low-dimensional systems like TMDCs, and to predict the level alignment at hybrid interfaces \citep{Neaton2006, Thygesen2009, Egger2015, Fu2016, Liu2019, Nabok2019}.~They enable the exploration of novel emergent phenomena and excitations as exemplified by hybrid charge-transfer excitons \citep{Bernardi2013, Turkina2018, Aggoune2020}, hyperbolicity \cite{Edalati-Boostan2020}, band-structure renormalization due to electron-phonon interactions \cite{Molina2015} or 2D polarons \cite{Kang2018, Caruso2021B}. 

In this work, we present a first-principles investigation of QPs and excitonic properties of organic-inorganic interfaces composed of a MoS$_2$ monolayer and the $\pi$-conjugated organic molecules pyrene (C$_{16}$H$_{10}$) and pyridine (C$_5$H$_5$N), respectively. We conduct $G_0W_0$-BSE calculations to obtain a predictive description of the quasiparticle (QP) band structures and the character and spatial distribution of (bound) excitons.~Our findings indicate that, despite the weakly-bound character of these materials, the interfacial level alignment and the optical properties are strongly affected by renormalization effects in both the inorganic and organic components, and high-level theory is needed to capture the underlying many-body effects.

\section{Computational details}

We use a slab model to describe the 2D unit cells of our hybrid systems. It consists of a  3$\times$3$\times$1 supercell of the pristine MoS$_2$ structure, {\it i.e.}, containing 9 molybdenum (Mo) and 18 sulfur (S) atoms. A vacuum region of 15 \AA~ and dipole corrections along the \textit{z}-direction are used to avoid spurious interactions between neighboring replica. The in-plane lattice parameter of 3.162~\AA\ is adopted from the optimized pristine MoS$_2$ unit cell. This result is in good agreement with former theoretical and experimental estimates \citep{Klots2014,Lembke2015,Molina2015,Qiu2016, Pisarra2021}.

Ground-state properties are calculated using DFT \citep{Hohenberg1964, Kohn1965} with the generalized gradient approximation in the Perdew-Burke-Ernzerhof (PBE) parameterization \citep{Perdew1996} for the exchange-correlation functional. The sampling of the Brillouin zone (BZ) is carried out with a homogeneous 15$\times$15$\times$1 Monkhorst-Pack \textbf{k}-point grid for the pristine MoS$_2$ and a 3$\times$3$\times$1 \textbf{k}-grid for the supercells. To account for vdW forces between substrate and molecules and inter-molecular interactions, we adopt the Tkatchenko-Scheffler (TS) method \citep{Tkatchenko2009}. 

\begin{figure}[t!]
\begin{center}
\includegraphics[width=0.8\columnwidth]{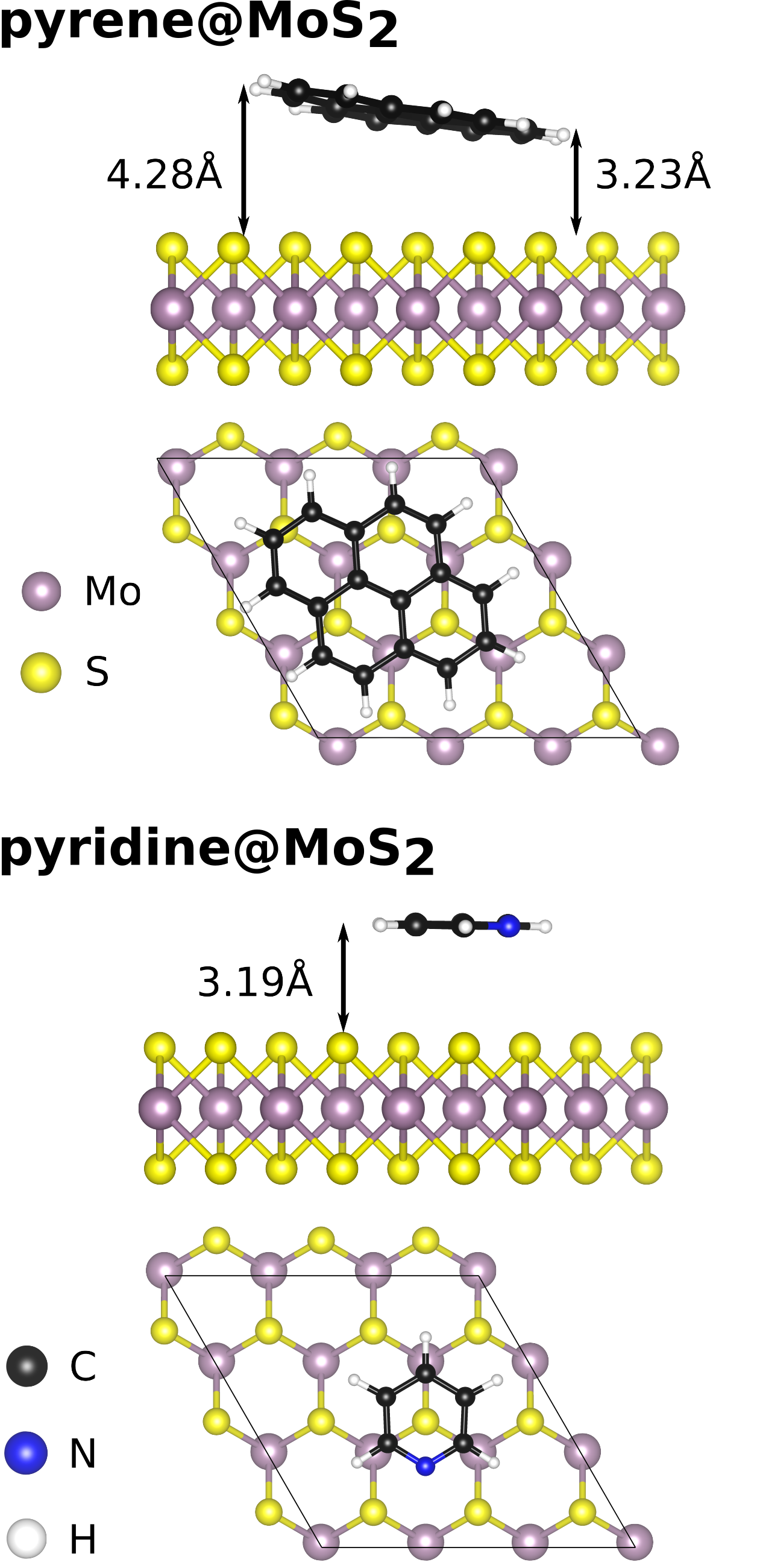}%
\caption{Side and top view of the optimized structures of pyrene@MoS$_2$ (top) and pyridine@MoS$_2$ (bottom) in a 3$\times$3$\times$1 MoS$_2$ supercell.}
\label{fig:structure}
\end{center}
\end{figure}
Structure optimizations are performed using the all-electron code FHI-aims \citep{Blum2009} by minimizing the amplitude of the interatomic forces below a threshold value of $10^{-3}$ eV \AA$^{-1}$. A TIER2 basis is used for the species of the molecules, and TIER1 and TIER1+dg for  Mo and S, respectively. 
The subsequent ground-state, $G_0W_0$, and BSE calculations are performed using \exciting~\citep{Gulans2014, Nabok2016, Vorwerk2019}, an all-electron full-potential code, implementing the family of linearized augmented planewave plus local orbitals (LAPW+LO) methods.~The muffin-tin (MT) spheres in the inorganic component are chosen to have equal radii of 2.2 bohr.~For the molecules, the corresponding radii are 0.8 bohr for  hydrogen (H) and 1.2 bohr for carbon (C) and nitrogen (N). We use a basis-set cutoff of G$_{\textrm{max}}$=4.375 bohr$^{-1}$.~QP energies are computed within the $G_{0}W_{0}$ approximation \citep{Hedin1965, Hybertsen1985} by solving the QP equation.~600 and 200 empty states are included to compute the frequency-dependent dielectric screening within the random-phase approximation in pristine MoS$_2$ and the hybrid systems, respectively. A 2D truncation of the Coulomb potential in the out-of-plane direction \textit{z} is employed \citep{Fu2016}. Band structure and density of states (DOS) are computed on dense meshes by using interpolation with maximally-localized Wannier functions \citep{Tillack2020}.

For the solution of the BSE~\cite{Hanke1980,Strinati1988, Rohlfing2000, Vorwerk2019} on top of the QP band structure, the screened Coulomb potential is computed using 100 empty bands. In the construction and diagonalization of the BSE Hamiltonian, 16 occupied and 14 unoccupied bands are included, and a 12$\times$12$\times$1 shifted $\textbf{k}$-point mesh is adopted for the hybrid systems. In the case of pristine MoS$_2$, we use 5 occupied, 7 unoccupied bands and a $\textbf{k}$-mesh of 18$\times$18$\times$1. These parameters ensure well-converged spectra within the energy window up to 6 eV. A Lorentzian broadening of 100 meV is applied to the spectra to mimic lifetime effects. Spin-orbit coupling is not considered but its possible impact is discussed in the Results section. 

Atomic structures and isosurfaces are visualized using the VESTA software \cite{Momma2011}. Input and output files of all calculations can be downloaded free of charge from the NOMAD Repository \cite{Draxl2019} by the following link: \href{https://dx.doi.org/10.17172/NOMAD/2021.11.02-4}{https://dx.doi.org/10.17172/NOMAD/2021.11.02-4}.

\section{Results and discussion}

\subsection{Structural properties}

For both pyrene and pyridine, we consider a starting configuration consisting of an up-right standing molecule at a vertical distance of about 3 \AA.  In case of pyridine, the nitrogen atom is pointing towards the monolayer. The internal coordinates of both systems are then relaxed. The side and top views of the optimized supercells are shown in Fig.~\ref{fig:structure}.~At equilibrium, the molecules are separated by about 3~\AA~from the monolayer. At this distance, no chemical bonding occurs, and the interactions between molecules and substrate is primarily due to vdW forces. The adsorption geometry of pyrene is tilted with an angle of 17$^{\circ}$. The shortest (largest) distance between the monolayer and the physisorbed molecules is 3.23~\AA\ (4.28~\AA), measured from the top of the substrate. This inclination is attributed to the weak interactions between neighboring molecules. Conversely, pyridine lies parallel to the ML plane, and the center of the aromatic ring is located above the S atom at an interlayer distance of 3.19~\AA~from the ML. 

To assess the structural stability of these geometries, we compute the adsorption energy as
\begin{equation}
E_a = E_{\mathrm {MoS_2/mol}} - E_{\mathrm{MoS_2}} - E_{\mathrm{mol}}.
\end{equation}
$E_{\mathrm{MoS_2}}$, $E_{\mathrm{mol}}$, and $E_{\mathrm {MoS_2/mol}}$ are the total energies of MoS$_2$, the molecule, and the hybrid heterostructure, respectively. For pyrene, we find an  adsorption energy of -2.13 eV, whereas for pyridine, we obtain -0.91 eV. We attribute the lower value of the latter to its smaller size, {\it i.e.} smaller surface contact area.

\begin{figure}[b!]
\begin{center}
\includegraphics[width=.8\columnwidth]{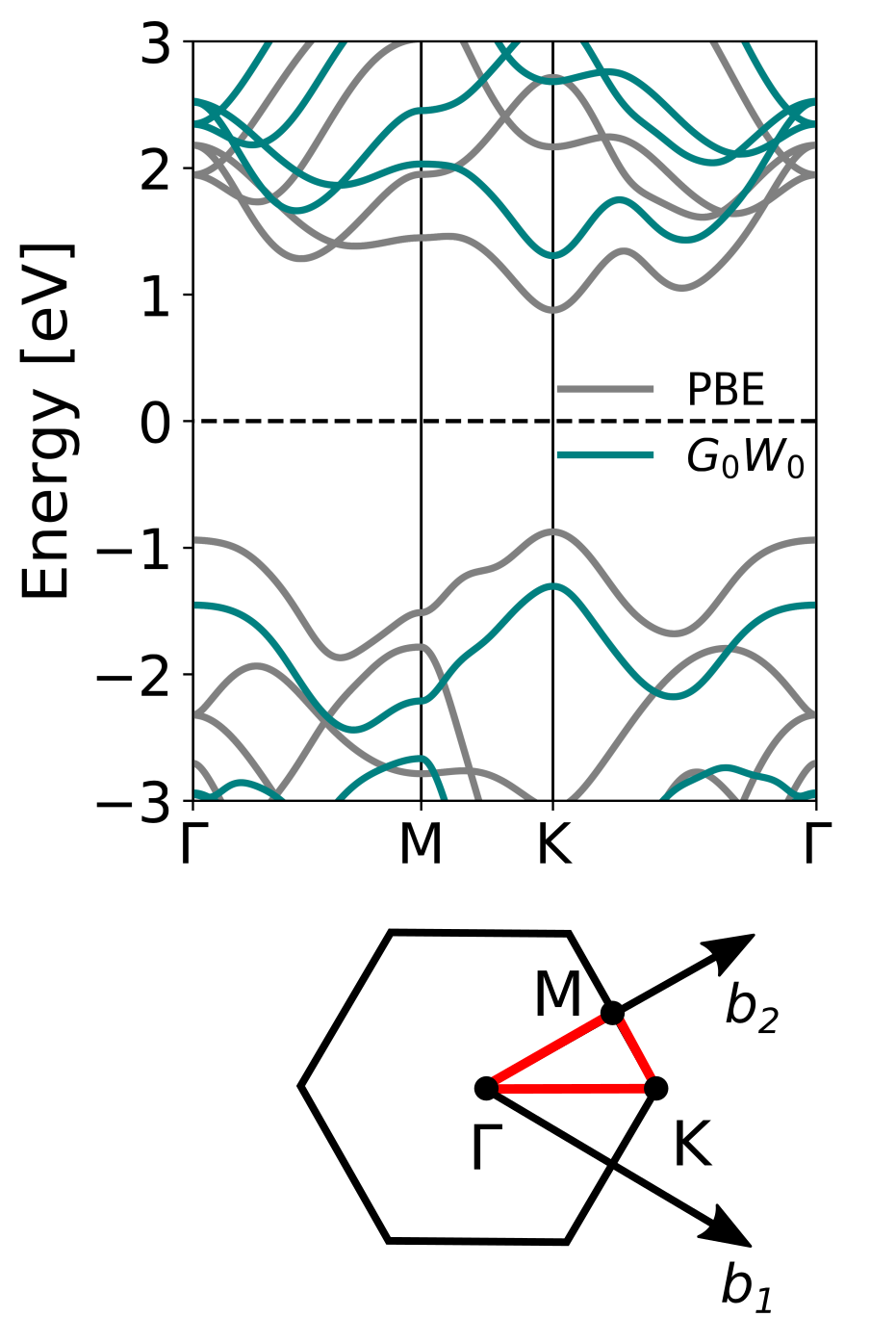}%
\caption{Band structure (top), as obtained from PBE (gray) and $G_0W_0$ (teal) of the pristine MoS$_2$ monolayer and Brillouin zone showing high-symmetry points (bottom).}
\label{fig:pristine}
\end{center}
\end{figure}
\subsection{Electronic structure}

The band structure of the pristine MoS$_2$ monolayer along the $\Gamma$-M-K-$\Gamma$ path as obtained from PBE+TS and $G_0W_0$ calculations is shown in the top of Fig.~\ref{fig:pristine}. The two-dimensional (2D) hexagonal Brillouin zone (BZ) is shown in the bottom of Fig.~\ref{fig:pristine}, indicating the main high-symmetry points. The direct Kohn-Sham band gap at K amounts to 1.7 eV, the QP gap to 2.6 eV, both values being in agreement with earlier calculations \citep{Molina2015, Qiu2016, Gonzalez2017, Caruso2021}. The QP gap also compares well with the estimated photocurrent gap of 2.5 eV \citep{Klots2014}. 

\begin{figure*}[ht!]
\begin{center}
\includegraphics[width=.99\textwidth]{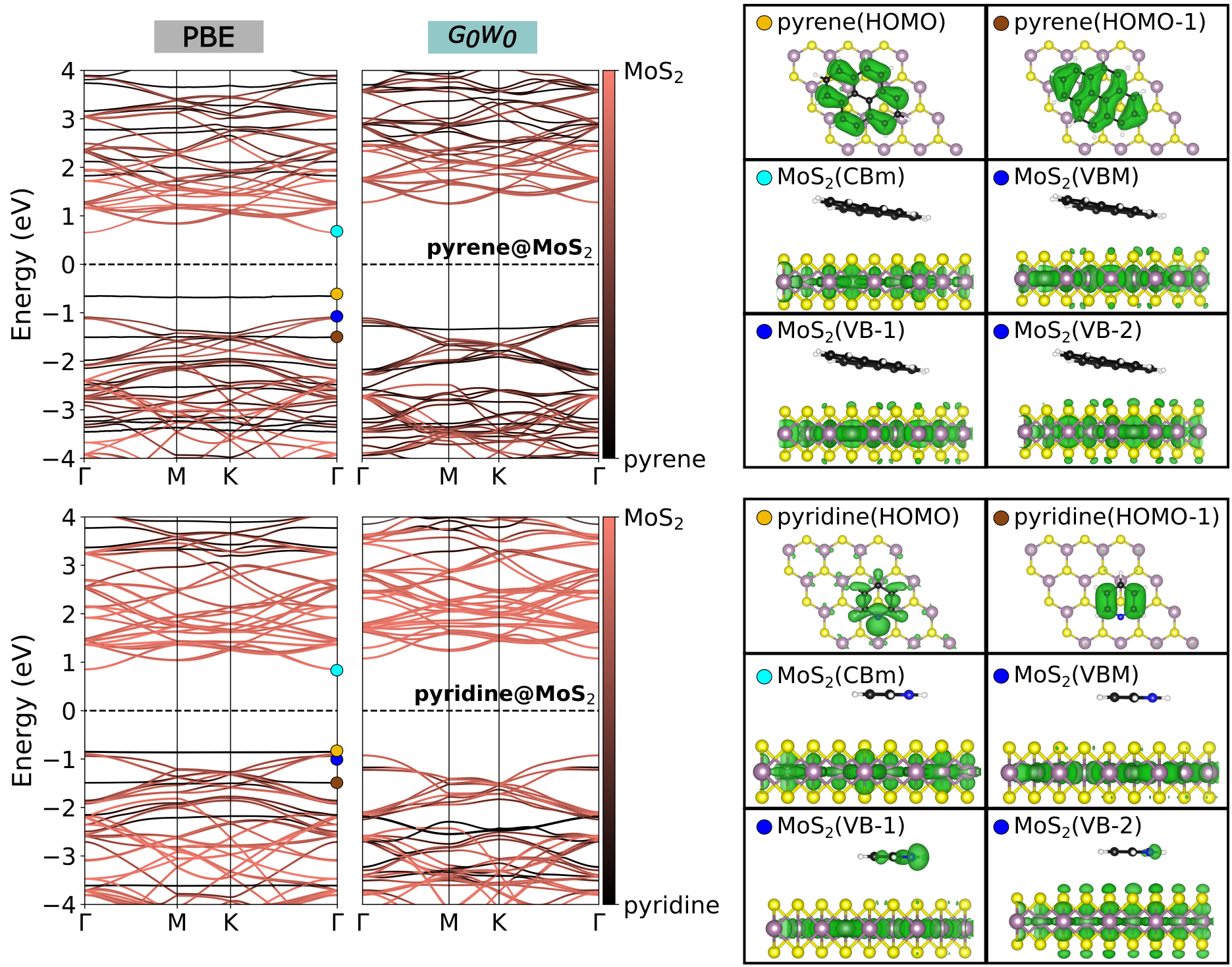}%
\caption{Band structures of pyrene@MoS$_2$ (top) and pyridine@MoS$_2$ (bottom), obtained from PBE and $G_0W_0$. The color code, going from red (MoS$_2$) to black (molecule) indicates the contribution for the pristine entities to the bands. The corresponding Kohn-Sham (KS) wavefunctions of selected bands at the $\Gamma$-point are shown on the right.}
\label{fig:electronic}
\end{center}
\end{figure*}

The PBE band structure of the hybrid heterostructures are depicted in Fig.~\ref{fig:electronic} (left panels).
Both materials exhibit direct gaps at the $\Gamma$ point of the supercell's BZ. Their appearance at $\Gamma$ point is a consequence of the band folding in the 3$\times$3$\times$1 supercell \citep{Valencia2017}. An analysis of the band characters indicate that such vdW-bound hybrid systems preserve the band structure of the pristine constituents on the PBE level to a large extent. However, careful inspection reveals that the electronic structure of the hybrid heterostructures is nevertheless affected by weak orbital hybridization and charge redistribution. For both systems, PBE yields an energy-level alignment of type II \citep{Turkina2018, Nabok2019}, where the HOMO of the molecule constitutes the highest occupied valence states, whereas the lowest-lying unoccupied bands arise from the conduction band of MoS$_2$. 

\begin{figure}[ht!]
\begin{center}
\includegraphics[width=.99\columnwidth]{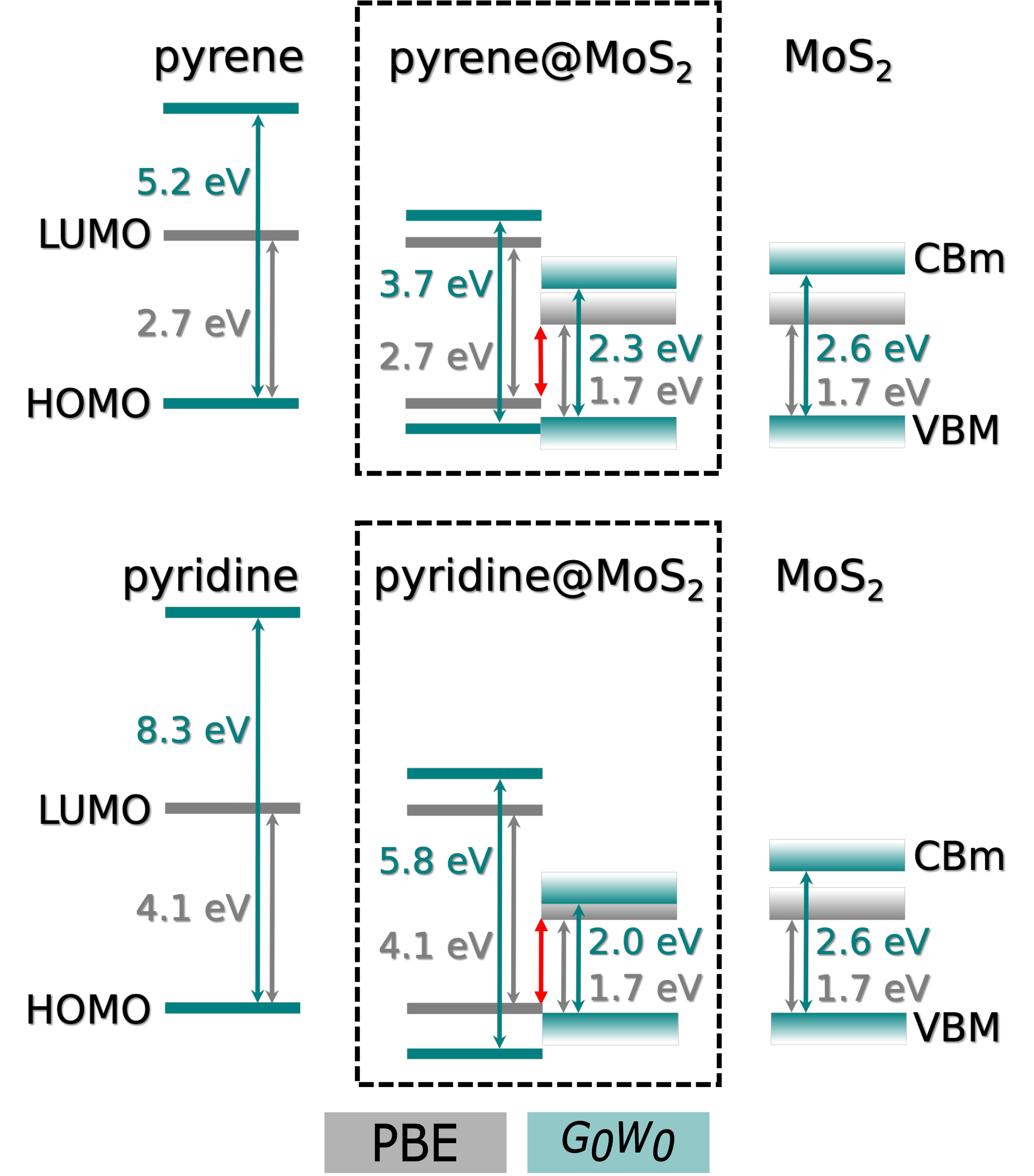}%
\caption{Energy-level alignment obtained from PBE (gray) and $G_0W_0$ (teal)  for pyrene@MoS$_2$ (top) and pyridine@MoS$_2$ (bottom). The middle panels show the respective hybrid material, on the left (right) the corresponding scheme is depicted for the molecular monolayer (MoS$_2$). The red arrows indicate the band gaps as obtained from DFT, corresponding to the (wrong) level alignment being type II compared to the type-I alignment obtained by $GW$ (teal arrows in the MoS$_2$ side).}
\label{fig:alignment}
\end{center}
\end{figure}
The adsorption of pyrene manifests itself through the emergence of a flat band around -0.6 eV in the valence region as shown in the top left panel of Fig.~\ref{fig:electronic}. Visualization of the corresponding wavefunction at the $\Gamma$ point (marked by a orange circle) in the right panel, confirms that this level corresponds to the HOMO of pyrene. The distance to the valence-band maximum (VBM) of MoS$_2$ amounts to 0.4 eV in pyrene@MoS$_2$, making hybridization of these states unlikely. Another flat energy level appears at around -1.5 eV. Inspecting the KS orbital (brown circle), it can be attributed to the HOMO-1 of pyrene. Both, HOMO and HOMO-1 are molecular orbitals with $\pi$ character \citep{Shirai2020}. The VBM, VB-1, and VB-2 are almost degenerate at $\Gamma$, with an energy of -1.1 eV. Visualization of the corresponding KS orbitals for the latter band edges reveals bands with a well defined MoS$_2$ character preserving the composition of Mo-4$d$ and S-3$p$ states (blue circles). Although hybridization of the molecular orbitals and MoS$_2$ bands exists, it turns out to be minimal. This can be explained by the fact that pyrene does not have a permanent dipole and,  thus, vdW forces dominate the formation of the hybrid heterostructure.

The situation is different in the case of pyridine adsorption. Here, the bands have mixed character as observed for the bands derived from MoS$_2$ VB-1/VB-2 states, (blue circles). Based on PBE, the HOMO of pyridine and the three uppermost valence bands (VBM, VB-1, and VB-2) of MoS$_2$ have a similar energy of -0.8 eV at $\Gamma$. Such a mixed character of top valence states has also been identified at the hybrid interface pyridine@ZnO~\citep{Turkina2018}, where it has been attributed to the high electronegativity of nitrogen and the permanent dipole of pyridine. Moreover, the HOMO of pyridine corresponds to an $sp^2$ orbital (Fig. \ref{fig:electronic} bottom right, orange circle), with contributions from carbon and nitrogen \citep{Marom2012}; and it also shows hybridization with Mo-4$d$ orbitals. We identify another flat band around -1.4 eV, corresponding to the HOMO-1 of pyridine (brown circle), which is a molecular orbital with  $\pi$ character. The PBE band gaps of MoS$_2$ (VBM-CBm) in both hybrid heterostructures coincide with the direct band gap in pristine MoS$_2$.
\begin{figure}[h!]
\begin{center}
\includegraphics[width=0.99\columnwidth]{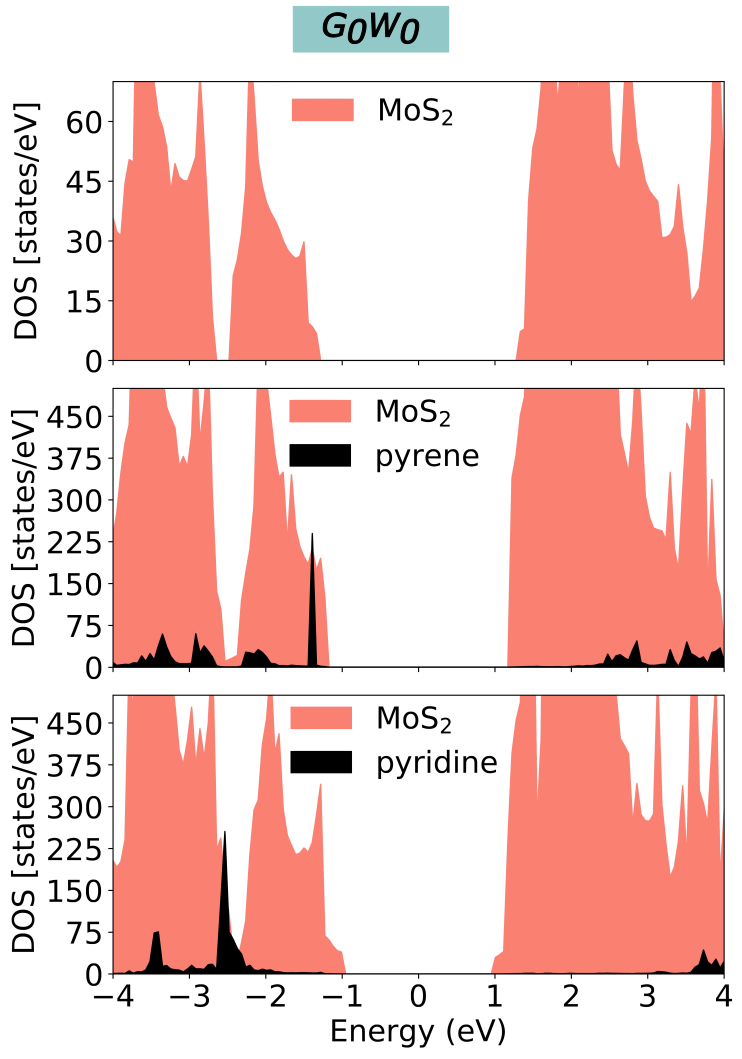}%
\caption{Density of states (DOS) in states per eV and unit cell calculated at the $G_0W_0$ level for  pristine MoS$_2$ (top), pyrene@MoS$_2$ (middle), and pyridine@MoS$_2$ (bottom panel).}
\label{fig:dos}
\end{center}
\end{figure}

It is well known that KS-DFT systematically underestimates band gaps of both molecules and solids \citep{Hybertsen1986, Jones1989, Onida2002, Cohen2012, Draxl2014}, typically to a different extent. Moreover, gap renormalization is expected due to dynamical screening effects, which are not accounted for by semi-local functionals like PBE \cite{Neaton2006, Puschnig2012,Fu2016,Marom2012, Marom2017}. To overcome these limitations, we compute the QP correction to the KS band structure employing the $G_0W_0$ approximation of MBPT. On the one hand, self-energy effects accounting for self-interaction correction increase the band gaps of both components, on the other hand, polarization effects induced by mutual interaction counteract the former. The results are depicted in Figs.~\ref{fig:electronic} and~\ref{fig:alignment}. In pyrene@MoS$_2$, the QP correction increases the gap of MoS$_2$ by 0.6 eV (from 1.7 to 2.3 eV), in pyridine@MoS$_2$ by 0.3 eV (from 1.7 to 2.0 eV). These values are significantly smaller than their counterpart in pristine MoS$_2$ in which the gap increases by 0.9 eV. Such reduction is to be expected as the molecular layer enhances the screening of the Coulomb interaction, \ie the $W$. The QP correction is also evident in the DOS, displayed for pristine MoS$_2$ (top) and the hybrid heterostructures (middle and bottom panels) in Fig.~\ref{fig:dos}. At first glance, one would anticipate that the presence of pyrene would impact MoS$_2$ more than pyridine, due to its larger spatial extension and much smaller HOMO-LUMO gap. However, we observe the opposite. This unexpected finding could be attributed to the fact that the pyridine molecules, exhibiting a permanent dipole (caused by the lone pair of electrons in nitrogen) represent a polarizable medium giving rise to non-local correlations in terms of an image-charge effect that affects the TMDC.

Also on the molecular side of the heterostructure, the electronic levels are renormalized by both effects, \ie the self-energy correction including image-charge effect, leading to a difference of the HOMO-LUMO gap relative to the value in the isolated molecular monolayer~\cite{Nabok2019}. The renormalization of the HOMO-LUMO gap of pyrene in the hybrid system amounts to 1.2 eV. A large amount arises from the pronounced self-interaction error of KS-DFT, being removed at the $G_0W_0$ level \citep{Marom2012, Caruso2016, Dauth2016}. However, the corresponding value of the isolated molecular monolayer, is even much larger, \ie 2.7 eV. In other words, the polarization-induced renormalization \citep{Neaton2006, Thygesen2009, Puschnig2012, Nabok2019} is as much as -1.5 eV.

Likewise, the HOMO-LUMO gap of pyridine in the heterostructure is renormalized by approximately 1.7 eV. In this case, the self-energy correction of the isolated molecular monolayer increases the gap approximately 4.2 eV,  while the polarization effect reduces the gap by about -2.5 eV. For both molecules, polarization effects have a smaller impact on the frontier molecular levels compared to self-interaction corrections. The larger polarization in pyridine@MoS$_2$ is again assigned to the permanent molecular dipole as well as the smaller interlayer distance of 3.19 \AA\ compared to the pyrene@MoS$_2$ interface.

Most important, the many-body effects on the energy levels of both the TMDC and the molecular layer lead to a qualitative change of the level alignment with respect to the result obtained by semilocal DFT (PBE). Most striking, in both material systems, the band alignment is changed to type I, with the HOMO of the respective molecule located below the MoS$_2$ VBM, VB-1, and VB-2. In the case of pyridine@MoS$_2$, the HOMO-derived band is located deep in the valence band region and intertwined with the HOMO-1 at an energy below -2 eV, as shown in the bottom left of Fig.~\ref{fig:electronic}. These findings re-emphasize the need for many-body corrections to reliably determine the level alignment of hybrid heterostructures \cite{Draxl2014}. While band alignment of type II usually favors charge transfer and exciton dissociation, type I, as found here, can also lead to exciton transfer and energy transfer processes. This interplay between various (mutual) effects on the individual components may be used as a tool to tailor the electronic properties of hybrid interfaces. Moreover, all of these features have an important impact on the optical properties of these systems, which are discussed in the next section.

\subsection{Optical excitations}
\begin{figure}[b!]
\begin{center}
\includegraphics[width=0.99\columnwidth]{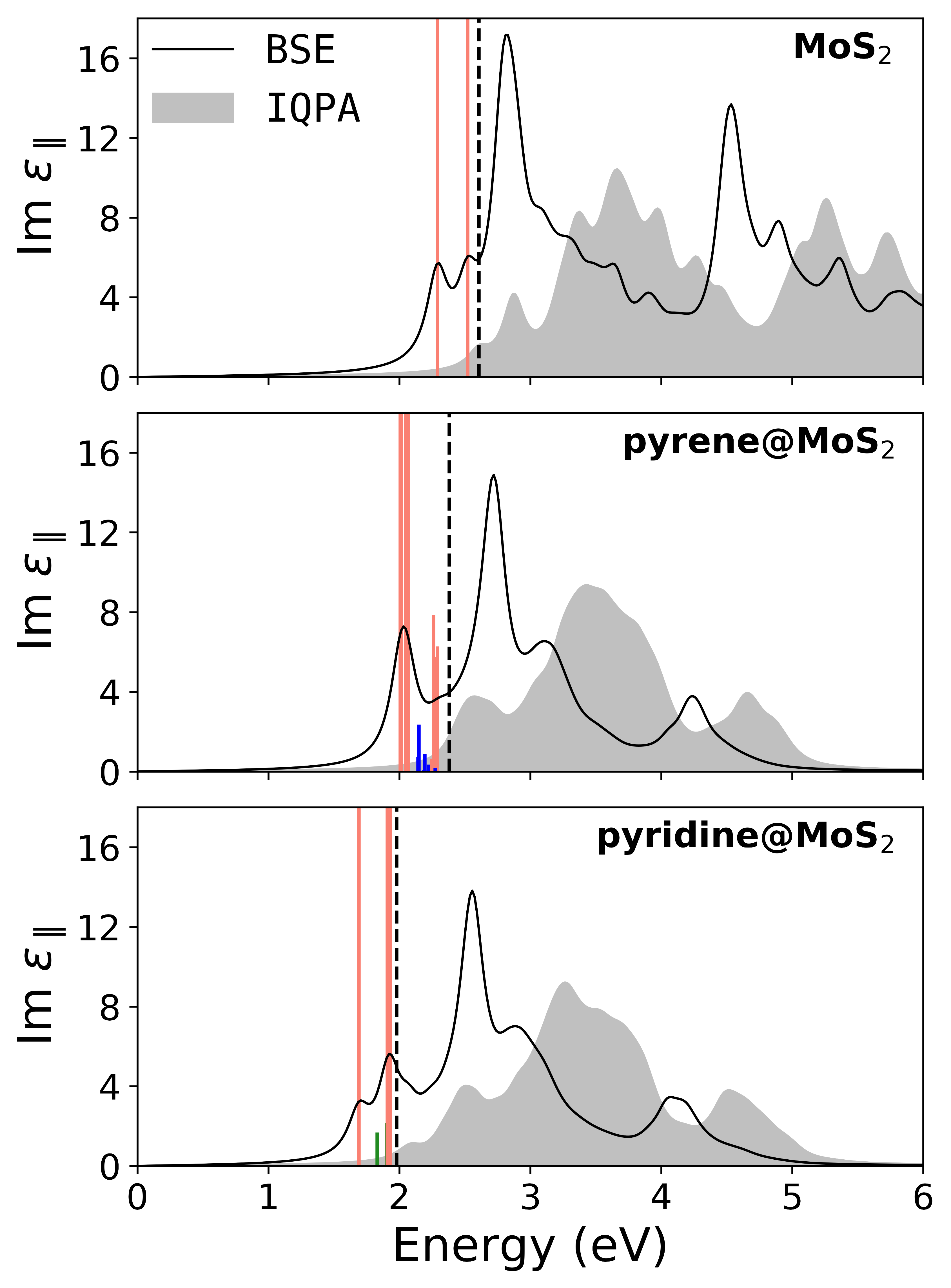}%
\caption{Dielectric function of pristine MoS$_2$ (top), pyrene@MoS$_2$ (middle), and pyridine@MoS$_2$ (bottom), averaged over the in-plane components, including (BSE, solid lines) and neglecting (IQPA, gray areas) excitonic effects. Red bars indicate the energetic position of MoS$_2$-derived excitons, blue bars those of (hybrid)~charge-transfer excitons, and green bars those of hybrid excitons. Their height corresponds to the oscillator strength. The black dashed lines mark the positions of the respective QP gap between the VBM and CBm of MoS$_2$.}
\label{fig:absorption}
\end{center}
\end{figure}
Based on the QP band structures, we compute optical properties of the heterostructures by the solution of the BSE. In Fig.~\ref{fig:absorption}, we display the imaginary  part of the frequency-dependent macroscopic dielectric function for pristine MoS$_2$ (top), pyrene@MoS$_2$ (middle panel) and pyridine@MoS$_2$ (bottom) as obtained by averaging over the in-plane components. The independent quasiparticle spectrum (IQPA) that ignores excitonic effects, is shown for comparison. 

In the following discussion, we focus on bound excitons, whose energies are located within the QP gap. These features result exclusively from electron-hole interactions and, thus, are absent in the IQPA. The absorption onset in pristine MoS$_2$ is located in the visible region at 2.29 eV. The binding energy of its first exciton amounts to 310 meV, in line with previous theoretical results (300-500 meV) \citep{Molina-Sanchez2013, Qiu2013}. In both interfaces, the lowest-energy exciton originates from MoS$_2$-derived transitions between the band edges around the VBM and the CBm. In pyrene@MoS$_2$ (pyridine@MoS$_2$), the corresponding excitonic peak is located at about 2.0 (1.7) eV, which is 380 (300) meV below its QP gap. Thus, compared to pristine MoS$_2$, the binding energy of the first exciton is by 70 meV higher in pyrene@MoS$_2$, but by 10 meV lower in pyridine@MoS$_2$.  

We classify the bound excitons according to the predominant character of the bands in which the involved electrons and holes are located. One can distinguish three different types of excitations: (i) MoS$_2$-like excitons, (ii) charge-transfer excitons, and (iii) hybrid excitons. A similar classification has been proposed for pyridine@ZnO interfaces \citep{Turkina2018}. Interestingly, in contrast to the latter work, here we find that both charge-transfer and hybrid excitons coexist in the visible region.
\begin{figure*}[th!]
\begin{center}
\includegraphics[width=.99\textwidth]{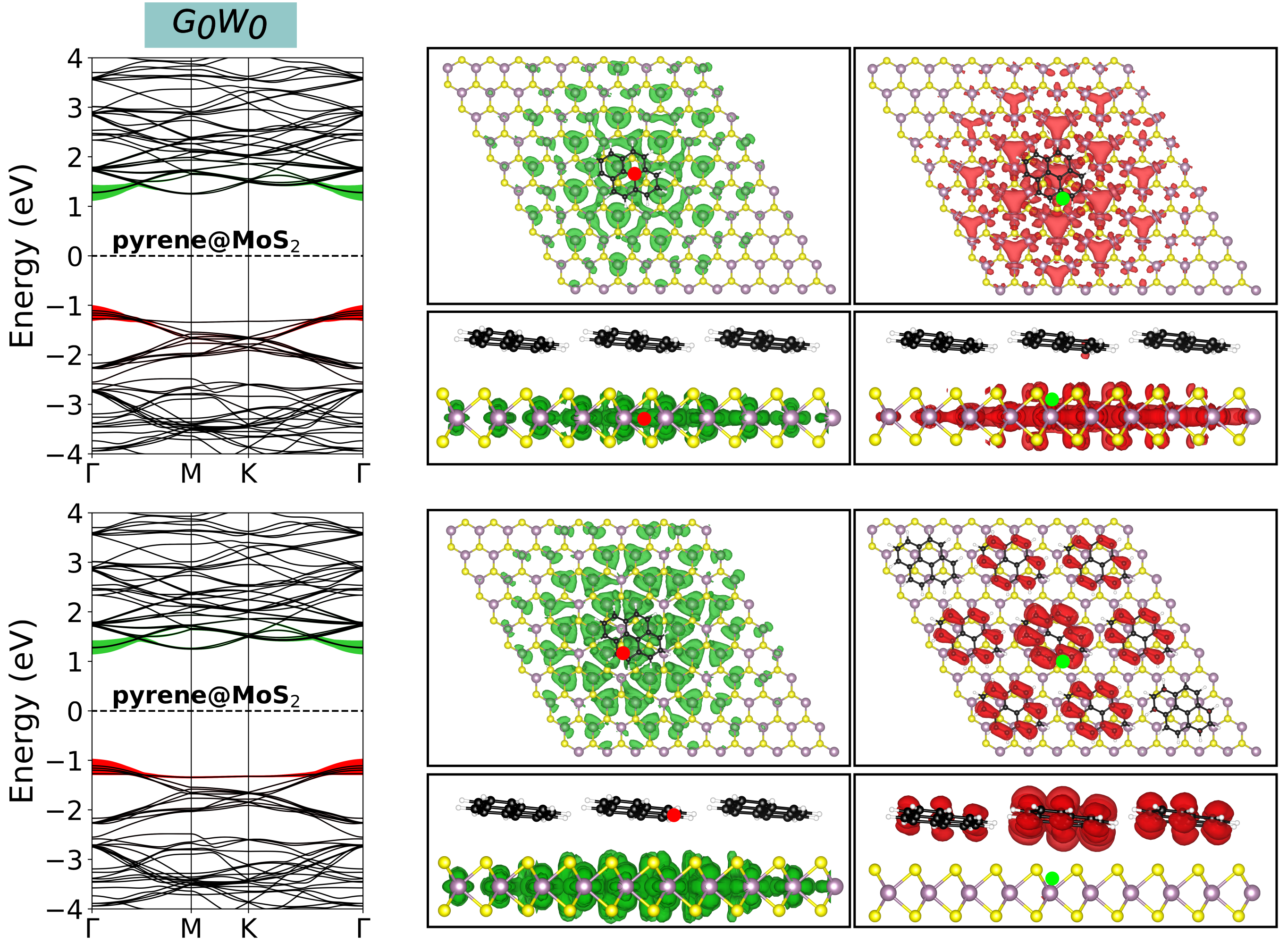}%
\caption{Reciprocal and real space representation of a MoS$_2$-like exciton (top) and charge-transfer exciton (bottom panels) in pyrene@MoS$_2$. The band structures on the left indicate the contributions of the involved states as red and green circles. Their sizes are proportional to the transition weight. The real space representations show the probability density of finding the hole of the e-h wavefunction given a fixed position of the electron (left panels) and vice versa (right panels). The electron (hole) probability distribution is depicted in green (red) with the corresponding hole (electron) position marked by the red (green) circles.}
\label{fig:exciton_pyrene}
\end{center}
\end{figure*}

\begin{figure*}[th!]
\begin{center}
\includegraphics[width=.99\textwidth]{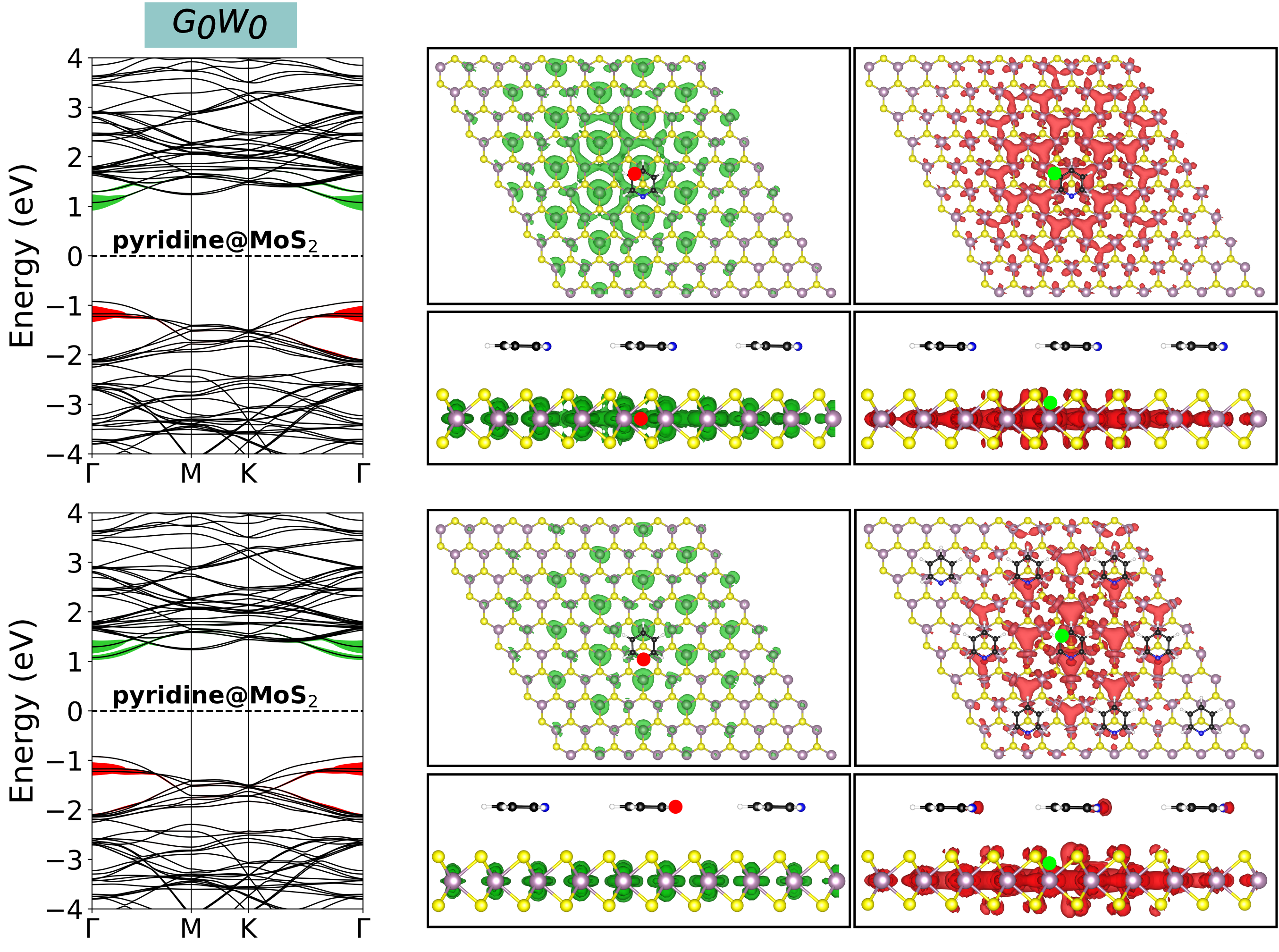}%
\caption{Reciprocal and real-space representation of an MoS$_2$-like exciton (top) and a hybrid exciton (bottom) in pyridine@MoS$_2$. The band structures on the left indicate the contributions of the involved states as red circles. Their sizes are proportional to the transition weight. The real-space representations show the probability density of finding the hole of the e-h wavefunction given a fixed position of the electron (left panels) and vice versa (right panels). The electron (hole) probability distribution is depicted in green (red) with the corresponding hole (electron) position marked by the red (green) circles.}
\label{fig:exciton_pyridine}
\end{center}
\end{figure*}

We proceed by picking one of each type of bound excitons for further analysis: (i) MoS$_2$-like excitons are characterized by the highest oscillator strengths (marked in red in Fig.~\ref{fig:absorption}) and involve transitions between the valence and conduction bands of MoS$_2$. As an example, we display in the top panels of Figs.~\ref{fig:exciton_pyrene} and \ref{fig:exciton_pyridine} the exciton wavefunctions corresponding to the first exciton of pyrene@MoS$_2$ and pyridine@MoS$_2$, respectively. In both, the real space plots on the right indicate moderately localized distributions of both the electron (green) and the hole (red) in this {\it e}-{\it h} pair indicative of a medium-sized exciton binding energy mentioned above (380 meV in pyrene@MoS$_2$, 310 meV in pyridine@MoS$_2$). Fixing the hole near a Mo atom (red dot in the left panel), the electron distribution reflects the Mo $d_{z^2}$ orbitals in the lowest MoS$_2$ conduction bands. While the electron distribution in both interfaces is similar, there is a distinction in the hole distribution. With the electron fixed near the Mo atom (green dot in the right panel), the mixed Mo-4$d$ and S-3$s$ character of the valence band becomes apparent. In both materials, the orbitals contributing to the hole distribution are the same but they exhibit a different orientation (compare the top right panels of Figs.~\ref{fig:exciton_pyrene} and~\ref{fig:exciton_pyridine}). We note that, overall, the MoS$_2$ band structure does not change significantly upon including spin-orbit coupling, the most prominent result being a splitting of the VBM at the K point by about 0.13. This leads to a splitting of the corresponding exciton peak in the optical spectra \citep{Marsili2021} where the lower (higher) one reflects excitations to the higher (lower) SOC-split band \citep{Vona2021}. Overall, the character of these excitons is the same. This finding can be transferred to these MoS$_2$-like excitations observed here.  

(ii) Charge-transfer excitons are characterized by small oscillator strengths (marked in blue in Fig.~\ref{fig:absorption}). The chosen representative of this type, at an energy of 2.15 eV, has a binding energy of 230 meV. This excitation involves transitions between the HOMO of pyrene and the lowest conduction band of MoS$_2$ as seen from the reciprocal-space analysis on the left in the bottom panel of Fig.~\ref{fig:exciton_pyrene}. The real space-plot shows that the exciton is more delocalized and has a different electronic distribution than the excitons of type (i).~A similar behavior has been found for other vdW materials, such as layered h-BN \citep{aggoune2018PRB}. When the hole is fixed on pyrene, the electron distribution mimics a projection of the molecular $\pi$ orbital onto the MoS$_2$ surface. With the electron fixed near a Mo atom, the hole distribution reflects the HOMO of pyrene (bottom right panel of Fig.~\ref{fig:exciton_pyrene}). This excitation, involving the molecular orbitals and the conduction bands of MoS$_2$ are allowed (and form bright excitons) due to their hybridization in the valence bands. 

(iii) Hybrid excitons involve transitions between predominantly hybridized bands and are thus typically characterized by small oscillator strengths As an example of a bound hybrid exciton, we select an excitation in pyridine@MoS$_2$ (marked in green in Fig.~\ref{fig:absorption}) with a binding energy of 200 meV at 1.8 eV, whose wavefunction is displayed in the bottom panel of Fig.~\ref{fig:absorption}. We see that the electron and hole distributions are delocalized over the substrate surface as depicted in the bottom panel of Fig.~\ref{fig:exciton_pyridine}. In the hole distribution, we see contributions from both the molecule and the substrate, a major difference to the previous type. As the hole is fixed here close to the nitrogen atom, a very delocalized electron distribution is found. The reciprocal-space analysis in Fig.~\ref{fig:exciton_pyridine} in the left panel shows transitions from the hybridized valence bands VB-1 and VB-2 to the hybridized conduction bands at the $\Gamma$ point.


\section*{Summary and conclusions}
In summary, we have presented first-principles calculations based on MBPT of the electronic and excitonic properties of hybrid van-der-Waals heterostructures. In particular, we have considered hybrid systems consisting of pyrene and pyridine physisorbed on a monolayer of MoS$_2$. $G_0W_0$ calculations reveal that many-body effects have a crucial impact on the type of level alignment at the interfaces. Our results indicate that the electronic and optical properties of MoS$_2$ can be tuned by both investigated molecules that influence the substrate in a different manner. In the case of pyrene, charge-transfer excitons are found while in the pyridine system, hybrid-like excitons are more prominent. In both materials, the renormalization of the molecular electronic levels plays an important role in the level alignment. The significantly more prominent polarization-induced effects in the pyridine@MoS$_2$ heterostructure are assigned to the permanent dipole of the molecule. The latter is also reflected in the higher hybridization of the valence bands. Overall, our findings indicate that hybrid systems consisting of low-dimensional semiconductors and organic $\pi$-conjugate molecules exhibit a rich variety of excitons, providing a promising playground for exploring many-body interactions and exciton physics.

\subsection*{Acknowledgment} 
This work was supported by the Deutscher Akademischer Austauschdienst (DAAD) and received funding from the Deutsche Forschungsgemeinschaft (DFG) – Projektnummer 182087777 – SFB 951. F.C. acknowledges funding by the DFG -- Projektnummer 443988403. I.G.O. acknowledges fruitful discussions with Wahib Aggoune and Sebastian Tillack.
%


\begin{thebibliography}{67}%
\makeatletter
\providecommand \@ifxundefined [1]{%
 \@ifx{#1\undefined}
}%
\providecommand \@ifnum [1]{%
 \ifnum #1\expandafter \@firstoftwo
 \else \expandafter \@secondoftwo
 \fi
}%
\providecommand \@ifx [1]{%
 \ifx #1\expandafter \@firstoftwo
 \else \expandafter \@secondoftwo
 \fi
}%
\providecommand \natexlab [1]{#1}%
\providecommand \enquote  [1]{``#1''}%
\providecommand \bibnamefont  [1]{#1}%
\providecommand \bibfnamefont [1]{#1}%
\providecommand \citenamefont [1]{#1}%
\providecommand \href@noop [0]{\@secondoftwo}%
\providecommand \href [0]{\begingroup \@sanitize@url \@href}%
\providecommand \@href[1]{\@@startlink{#1}\@@href}%
\providecommand \@@href[1]{\endgroup#1\@@endlink}%
\providecommand \@sanitize@url [0]{\catcode `\\12\catcode `\$12\catcode
  `\&12\catcode `\#12\catcode `\^12\catcode `\_12\catcode `\%12\relax}%
\providecommand \@@startlink[1]{}%
\providecommand \@@endlink[0]{}%
\providecommand \url  [0]{\begingroup\@sanitize@url \@url }%
\providecommand \@url [1]{\endgroup\@href {#1}{\urlprefix }}%
\providecommand \urlprefix  [0]{URL }%
\providecommand \Eprint [0]{\href }%
\providecommand \doibase [0]{http://dx.doi.org/}%
\providecommand \selectlanguage [0]{\@gobble}%
\providecommand \bibinfo  [0]{\@secondoftwo}%
\providecommand \bibfield  [0]{\@secondoftwo}%
\providecommand \translation [1]{[#1]}%
\providecommand \BibitemOpen [0]{}%
\providecommand \bibitemStop [0]{}%
\providecommand \bibitemNoStop [0]{.\EOS\space}%
\providecommand \EOS [0]{\spacefactor3000\relax}%
\providecommand \BibitemShut  [1]{\csname bibitem#1\endcsname}%
\let\auto@bib@innerbib\@empty
\bibitem [{\citenamefont {Klots}\ \emph {et~al.}(2014)\citenamefont {Klots},
  \citenamefont {Newaz}, \citenamefont {Wang}, \citenamefont {Prasai},
  \citenamefont {Krzyzanowska}, \citenamefont {Lin}, \citenamefont {Caudel},
  \citenamefont {Ghimire}, \citenamefont {Yan}, \citenamefont {Ivanov},
  \citenamefont {Velizhanin}, \citenamefont {Burger}, \citenamefont {Mandrus},
  \citenamefont {Tolk}, \citenamefont {Pantelides},\ and\ \citenamefont
  {Bolotin}}]{Klots2014}%
  \BibitemOpen
  \bibfield  {author} {\bibinfo {author} {\bibfnamefont {A.~R.}\ \bibnamefont
  {Klots}}, \bibinfo {author} {\bibfnamefont {A.~K.}\ \bibnamefont {Newaz}},
  \bibinfo {author} {\bibfnamefont {B.}~\bibnamefont {Wang}}, \bibinfo {author}
  {\bibfnamefont {D.}~\bibnamefont {Prasai}}, \bibinfo {author} {\bibfnamefont
  {H.}~\bibnamefont {Krzyzanowska}}, \bibinfo {author} {\bibfnamefont
  {J.}~\bibnamefont {Lin}}, \bibinfo {author} {\bibfnamefont {D.}~\bibnamefont
  {Caudel}}, \bibinfo {author} {\bibfnamefont {N.~J.}\ \bibnamefont {Ghimire}},
  \bibinfo {author} {\bibfnamefont {J.}~\bibnamefont {Yan}}, \bibinfo {author}
  {\bibfnamefont {B.~L.}\ \bibnamefont {Ivanov}}, \bibinfo {author}
  {\bibfnamefont {K.~A.}\ \bibnamefont {Velizhanin}}, \bibinfo {author}
  {\bibfnamefont {A.}~\bibnamefont {Burger}}, \bibinfo {author} {\bibfnamefont
  {D.~G.}\ \bibnamefont {Mandrus}}, \bibinfo {author} {\bibfnamefont {N.~H.}\
  \bibnamefont {Tolk}}, \bibinfo {author} {\bibfnamefont {S.~T.}\ \bibnamefont
  {Pantelides}}, \ and\ \bibinfo {author} {\bibfnamefont {K.~I.}\ \bibnamefont
  {Bolotin}},\ }\href {\doibase 10.1038/srep06608} {\bibfield  {journal}
  {\bibinfo  {journal} {Sci.~Rep.~}\ }\textbf {\bibinfo {volume} {4}},\
  \bibinfo {pages} {6608} (\bibinfo {year} {2014})}\BibitemShut {NoStop}%
\bibitem [{\citenamefont {Lembke}\ \emph {et~al.}(2015)\citenamefont {Lembke},
  \citenamefont {Bertolazzi},\ and\ \citenamefont {Kis}}]{Lembke2015}%
  \BibitemOpen
  \bibfield  {author} {\bibinfo {author} {\bibfnamefont {D.}~\bibnamefont
  {Lembke}}, \bibinfo {author} {\bibfnamefont {S.}~\bibnamefont {Bertolazzi}},
  \ and\ \bibinfo {author} {\bibfnamefont {A.}~\bibnamefont {Kis}},\ }\href
  {\doibase 10.1021/ar500274q} {\bibfield  {journal} {\bibinfo  {journal}
  {Acc.~Chem.~Res.~}\ }\textbf {\bibinfo {volume} {48}},\ \bibinfo {pages}
  {100} (\bibinfo {year} {2015})}\BibitemShut {NoStop}%
\bibitem [{\citenamefont {Manzeli}\ \emph {et~al.}(2017)\citenamefont
  {Manzeli}, \citenamefont {Ovchinnikov}, \citenamefont {Pasquier},
  \citenamefont {Yazyev},\ and\ \citenamefont {Kis}}]{Manzeli2017}%
  \BibitemOpen
  \bibfield  {author} {\bibinfo {author} {\bibfnamefont {S.}~\bibnamefont
  {Manzeli}}, \bibinfo {author} {\bibfnamefont {D.}~\bibnamefont
  {Ovchinnikov}}, \bibinfo {author} {\bibfnamefont {D.}~\bibnamefont
  {Pasquier}}, \bibinfo {author} {\bibfnamefont {O.~V.}\ \bibnamefont
  {Yazyev}}, \ and\ \bibinfo {author} {\bibfnamefont {A.}~\bibnamefont {Kis}},\
  }\href {\doibase 10.1038/natrevmats.2017.33} {\bibfield  {journal} {\bibinfo
  {journal} {Nat.~Rev.~Mat.~}\ }\textbf {\bibinfo {volume} {2}},\ \bibinfo
  {pages} {17033} (\bibinfo {year} {2017})}\BibitemShut {NoStop}%
\bibitem [{\citenamefont {Kwon}\ \emph {et~al.}(2019)\citenamefont {Kwon},
  \citenamefont {Garg}, \citenamefont {Park}, \citenamefont {Jeong},
  \citenamefont {Yu}, \citenamefont {Kim}, \citenamefont {Kung},\ and\
  \citenamefont {Im}}]{Kwon2019}%
  \BibitemOpen
  \bibfield  {author} {\bibinfo {author} {\bibfnamefont {H.}~\bibnamefont
  {Kwon}}, \bibinfo {author} {\bibfnamefont {S.}~\bibnamefont {Garg}}, \bibinfo
  {author} {\bibfnamefont {J.~H.}\ \bibnamefont {Park}}, \bibinfo {author}
  {\bibfnamefont {Y.}~\bibnamefont {Jeong}}, \bibinfo {author} {\bibfnamefont
  {S.}~\bibnamefont {Yu}}, \bibinfo {author} {\bibfnamefont {S.~M.}\
  \bibnamefont {Kim}}, \bibinfo {author} {\bibfnamefont {P.}~\bibnamefont
  {Kung}}, \ and\ \bibinfo {author} {\bibfnamefont {S.}~\bibnamefont {Im}},\
  }\href {\doibase 10.1038/s41699-019-0091-9} {\bibfield  {journal} {\bibinfo
  {journal} {npj 2D Mater. Appl.}\ }\textbf {\bibinfo {volume} {3}},\ \bibinfo
  {pages} {9} (\bibinfo {year} {2019})}\BibitemShut {NoStop}%
\bibitem [{\citenamefont {Romaner}\ \emph {et~al.}(2008)\citenamefont
  {Romaner}, \citenamefont {Heimel}, \citenamefont {Ambrosch-Draxl},\ and\
  \citenamefont {Zojer}}]{Romaner2008}%
  \BibitemOpen
  \bibfield  {author} {\bibinfo {author} {\bibfnamefont {L.}~\bibnamefont
  {Romaner}}, \bibinfo {author} {\bibfnamefont {G.}~\bibnamefont {Heimel}},
  \bibinfo {author} {\bibfnamefont {C.}~\bibnamefont {Ambrosch-Draxl}}, \ and\
  \bibinfo {author} {\bibfnamefont {E.}~\bibnamefont {Zojer}},\ }\href
  {\doibase 10.1002/adfm.200800876} {\bibfield  {journal} {\bibinfo  {journal}
  {Adv.~Funct.~Mater.~}\ }\textbf {\bibinfo {volume} {18}},\ \bibinfo {pages}
  {3999} (\bibinfo {year} {2008})}\BibitemShut {NoStop}%
\bibitem [{\citenamefont {Garcia-Lastra}\ \emph {et~al.}(2009)\citenamefont
  {Garcia-Lastra}, \citenamefont {Rostgaard}, \citenamefont {Rubio},\ and\
  \citenamefont {Thygesen}}]{Garcia-Lastra2009}%
  \BibitemOpen
  \bibfield  {author} {\bibinfo {author} {\bibfnamefont {J.~M.}\ \bibnamefont
  {Garcia-Lastra}}, \bibinfo {author} {\bibfnamefont {C.}~\bibnamefont
  {Rostgaard}}, \bibinfo {author} {\bibfnamefont {A.}~\bibnamefont {Rubio}}, \
  and\ \bibinfo {author} {\bibfnamefont {K.~S.}\ \bibnamefont {Thygesen}},\
  }\href {\doibase 10.1103/PhysRevB.80.245427} {\bibfield  {journal} {\bibinfo
  {journal} {Phys.~Rev.~B}\ }\textbf {\bibinfo {volume} {80}},\ \bibinfo
  {pages} {245427} (\bibinfo {year} {2009})}\BibitemShut {NoStop}%
\bibitem [{\citenamefont {Puschnig}\ \emph {et~al.}(2012)\citenamefont
  {Puschnig}, \citenamefont {Amiri},\ and\ \citenamefont
  {Draxl}}]{Puschnig2012}%
  \BibitemOpen
  \bibfield  {author} {\bibinfo {author} {\bibfnamefont {P.}~\bibnamefont
  {Puschnig}}, \bibinfo {author} {\bibfnamefont {P.}~\bibnamefont {Amiri}}, \
  and\ \bibinfo {author} {\bibfnamefont {C.}~\bibnamefont {Draxl}},\ }\href
  {\doibase 10.1103/PhysRevB.86.085107} {\bibfield  {journal} {\bibinfo
  {journal} {Phys.~Rev.~B}\ }\textbf {\bibinfo {volume} {86}},\ \bibinfo
  {pages} {085107} (\bibinfo {year} {2012})},\ \Eprint
  {http://arxiv.org/abs/1204.5289} {1204.5289} \BibitemShut {NoStop}%
\bibitem [{\citenamefont {Schlesinger}\ \emph {et~al.}(2015)\citenamefont
  {Schlesinger}, \citenamefont {Bianchi}, \citenamefont {Blumstengel},
  \citenamefont {Christodoulou}, \citenamefont {Ovsyannikov}, \citenamefont
  {Kobin}, \citenamefont {Moudgil}, \citenamefont {Barlow}, \citenamefont
  {Hecht}, \citenamefont {Marder}, \citenamefont {Henneberger},\ and\
  \citenamefont {Koch}}]{Schlesinger2015}%
  \BibitemOpen
  \bibfield  {author} {\bibinfo {author} {\bibfnamefont {R.}~\bibnamefont
  {Schlesinger}}, \bibinfo {author} {\bibfnamefont {F.}~\bibnamefont
  {Bianchi}}, \bibinfo {author} {\bibfnamefont {S.}~\bibnamefont
  {Blumstengel}}, \bibinfo {author} {\bibfnamefont {C.}~\bibnamefont
  {Christodoulou}}, \bibinfo {author} {\bibfnamefont {R.}~\bibnamefont
  {Ovsyannikov}}, \bibinfo {author} {\bibfnamefont {B.}~\bibnamefont {Kobin}},
  \bibinfo {author} {\bibfnamefont {K.}~\bibnamefont {Moudgil}}, \bibinfo
  {author} {\bibfnamefont {S.}~\bibnamefont {Barlow}}, \bibinfo {author}
  {\bibfnamefont {S.}~\bibnamefont {Hecht}}, \bibinfo {author} {\bibfnamefont
  {S.~R.}\ \bibnamefont {Marder}}, \bibinfo {author} {\bibfnamefont
  {F.}~\bibnamefont {Henneberger}}, \ and\ \bibinfo {author} {\bibfnamefont
  {N.}~\bibnamefont {Koch}},\ }\href {\doibase 10.1038/ncomms7754} {\bibfield
  {journal} {\bibinfo  {journal} {Nature~Commun.}\ }\textbf {\bibinfo {volume}
  {6}},\ \bibinfo {pages} {6754} (\bibinfo {year} {2015})}\BibitemShut
  {NoStop}%
\bibitem [{\citenamefont {Otero}\ \emph {et~al.}(2017)\citenamefont {Otero},
  \citenamefont {{V{\'{a}}zquez de Parga}},\ and\ \citenamefont
  {Gallego}}]{Otero2017}%
  \BibitemOpen
  \bibfield  {author} {\bibinfo {author} {\bibfnamefont {R.}~\bibnamefont
  {Otero}}, \bibinfo {author} {\bibfnamefont {A.~L.}\ \bibnamefont
  {{V{\'{a}}zquez de Parga}}}, \ and\ \bibinfo {author} {\bibfnamefont {J.~M.}\
  \bibnamefont {Gallego}},\ }\href {\doibase 10.1016/j.surfrep.2017.03.001}
  {\bibfield  {journal} {\bibinfo  {journal} {Surf.~Sci.~Rep.~}\ }\textbf
  {\bibinfo {volume} {72}},\ \bibinfo {pages} {105} (\bibinfo {year}
  {2017})}\BibitemShut {NoStop}%
\bibitem [{\citenamefont {Molina-S{\'{a}}nchez}\ \emph
  {et~al.}(2013)\citenamefont {Molina-S{\'{a}}nchez}, \citenamefont {Sangalli},
  \citenamefont {Hummer}, \citenamefont {Marini},\ and\ \citenamefont
  {Wirtz}}]{Molina-Sanchez2013}%
  \BibitemOpen
  \bibfield  {author} {\bibinfo {author} {\bibfnamefont {A.}~\bibnamefont
  {Molina-S{\'{a}}nchez}}, \bibinfo {author} {\bibfnamefont {D.}~\bibnamefont
  {Sangalli}}, \bibinfo {author} {\bibfnamefont {K.}~\bibnamefont {Hummer}},
  \bibinfo {author} {\bibfnamefont {A.}~\bibnamefont {Marini}}, \ and\ \bibinfo
  {author} {\bibfnamefont {L.}~\bibnamefont {Wirtz}},\ }\href {\doibase
  10.1103/PhysRevB.88.045412} {\bibfield  {journal} {\bibinfo  {journal}
  {Phys.~Rev.~B}\ }\textbf {\bibinfo {volume} {88}},\ \bibinfo {pages} {045412}
  (\bibinfo {year} {2013})}\BibitemShut {NoStop}%
\bibitem [{\citenamefont {Molina-S{\'{a}}nchez}\ \emph
  {et~al.}(2015)\citenamefont {Molina-S{\'{a}}nchez}, \citenamefont {Hummer},\
  and\ \citenamefont {Wirtz}}]{Molina2015}%
  \BibitemOpen
  \bibfield  {author} {\bibinfo {author} {\bibfnamefont {A.}~\bibnamefont
  {Molina-S{\'{a}}nchez}}, \bibinfo {author} {\bibfnamefont {K.}~\bibnamefont
  {Hummer}}, \ and\ \bibinfo {author} {\bibfnamefont {L.}~\bibnamefont
  {Wirtz}},\ }\href {\doibase 10.1016/j.surfrep.2015.10.001} {\bibfield
  {journal} {\bibinfo  {journal} {Surf.~Sci.~Rep.~}\ }\textbf {\bibinfo
  {volume} {70}},\ \bibinfo {pages} {554} (\bibinfo {year} {2015})}\BibitemShut
  {NoStop}%
\bibitem [{\citenamefont {Qiu}\ \emph {et~al.}(2013)\citenamefont {Qiu},
  \citenamefont {{Da Jornada}},\ and\ \citenamefont {Louie}}]{Qiu2013}%
  \BibitemOpen
  \bibfield  {author} {\bibinfo {author} {\bibfnamefont {D.~Y.}\ \bibnamefont
  {Qiu}}, \bibinfo {author} {\bibfnamefont {F.~H.}\ \bibnamefont {{Da
  Jornada}}}, \ and\ \bibinfo {author} {\bibfnamefont {S.~G.}\ \bibnamefont
  {Louie}},\ }\href {\doibase 10.1103/PhysRevLett.111.216805} {\bibfield
  {journal} {\bibinfo  {journal} {Phys.~Rev.~Lett.~}\ }\textbf {\bibinfo
  {volume} {111}},\ \bibinfo {pages} {216805} (\bibinfo {year}
  {2013})}\BibitemShut {NoStop}%
\bibitem [{\citenamefont {Qiu}\ \emph {et~al.}(2016)\citenamefont {Qiu},
  \citenamefont {{Da Jornada}},\ and\ \citenamefont {Louie}}]{Qiu2016}%
  \BibitemOpen
  \bibfield  {author} {\bibinfo {author} {\bibfnamefont {D.~Y.}\ \bibnamefont
  {Qiu}}, \bibinfo {author} {\bibfnamefont {F.~H.}\ \bibnamefont {{Da
  Jornada}}}, \ and\ \bibinfo {author} {\bibfnamefont {S.~G.}\ \bibnamefont
  {Louie}},\ }\href {\doibase 10.1103/PhysRevB.93.235435} {\bibfield  {journal}
  {\bibinfo  {journal} {Phys.~Rev.~B}\ }\textbf {\bibinfo {volume} {93}},\
  \bibinfo {pages} {235435} (\bibinfo {year} {2016})}\BibitemShut {NoStop}%
\bibitem [{\citenamefont {Hagara}\ \emph {et~al.}(2020)\citenamefont {Hagara},
  \citenamefont {Mrkyvkova}, \citenamefont {N{\'{a}}da{\v{z}}dy}, \citenamefont
  {Hodas}, \citenamefont {Bod{\'{i}}k}, \citenamefont {Jergel}, \citenamefont
  {Majkov{\'{a}}}, \citenamefont {Tok{\'{a}}r}, \citenamefont {Hut{\'{a}}r},
  \citenamefont {Sojkov{\'{a}}}, \citenamefont {Chumakov}, \citenamefont
  {Konovalov}, \citenamefont {Pandit}, \citenamefont {Roth}, \citenamefont
  {Hinderhofer}, \citenamefont {Hulman}, \citenamefont {Siffalovic},\ and\
  \citenamefont {Schreiber}}]{Hagara2020}%
  \BibitemOpen
  \bibfield  {author} {\bibinfo {author} {\bibfnamefont {J.}~\bibnamefont
  {Hagara}}, \bibinfo {author} {\bibfnamefont {N.}~\bibnamefont {Mrkyvkova}},
  \bibinfo {author} {\bibfnamefont {P.}~\bibnamefont {N{\'{a}}da{\v{z}}dy}},
  \bibinfo {author} {\bibfnamefont {M.}~\bibnamefont {Hodas}}, \bibinfo
  {author} {\bibfnamefont {M.}~\bibnamefont {Bod{\'{i}}k}}, \bibinfo {author}
  {\bibfnamefont {M.}~\bibnamefont {Jergel}}, \bibinfo {author} {\bibfnamefont
  {E.}~\bibnamefont {Majkov{\'{a}}}}, \bibinfo {author} {\bibfnamefont
  {K.}~\bibnamefont {Tok{\'{a}}r}}, \bibinfo {author} {\bibfnamefont
  {P.}~\bibnamefont {Hut{\'{a}}r}}, \bibinfo {author} {\bibfnamefont
  {M.}~\bibnamefont {Sojkov{\'{a}}}}, \bibinfo {author} {\bibfnamefont
  {A.}~\bibnamefont {Chumakov}}, \bibinfo {author} {\bibfnamefont
  {O.}~\bibnamefont {Konovalov}}, \bibinfo {author} {\bibfnamefont
  {P.}~\bibnamefont {Pandit}}, \bibinfo {author} {\bibfnamefont
  {S.}~\bibnamefont {Roth}}, \bibinfo {author} {\bibfnamefont {A.}~\bibnamefont
  {Hinderhofer}}, \bibinfo {author} {\bibfnamefont {M.}~\bibnamefont {Hulman}},
  \bibinfo {author} {\bibfnamefont {P.}~\bibnamefont {Siffalovic}}, \ and\
  \bibinfo {author} {\bibfnamefont {F.}~\bibnamefont {Schreiber}},\ }\href
  {\doibase 10.1039/c9cp05728e} {\bibfield  {journal} {\bibinfo  {journal}
  {Phys.~Chem.~Chem.~Phys.~}\ }\textbf {\bibinfo {volume} {22}},\ \bibinfo
  {pages} {3097} (\bibinfo {year} {2020})}\BibitemShut {NoStop}%
\bibitem [{\citenamefont {Caruso}(2021)}]{Caruso2021}%
  \BibitemOpen
  \bibfield  {author} {\bibinfo {author} {\bibfnamefont {F.}~\bibnamefont
  {Caruso}},\ }\href {\doibase 10.1021/acs.jpclett.0c03616} {\bibfield
  {journal} {\bibinfo  {journal} {J.~Phys.~Chem.~Lett.}\ }\textbf {\bibinfo
  {volume} {12}},\ \bibinfo {pages} {1734} (\bibinfo {year}
  {2021})}\BibitemShut {NoStop}%
\bibitem [{\citenamefont {Liu}\ \emph {et~al.}(2017)\citenamefont {Liu},
  \citenamefont {Gu}, \citenamefont {Ding}, \citenamefont {Fan}, \citenamefont
  {Hu}, \citenamefont {Tseng}, \citenamefont {Lee}, \citenamefont {Menon},\
  and\ \citenamefont {Forrest}}]{Liu2017}%
  \BibitemOpen
  \bibfield  {author} {\bibinfo {author} {\bibfnamefont {X.}~\bibnamefont
  {Liu}}, \bibinfo {author} {\bibfnamefont {J.}~\bibnamefont {Gu}}, \bibinfo
  {author} {\bibfnamefont {K.}~\bibnamefont {Ding}}, \bibinfo {author}
  {\bibfnamefont {D.}~\bibnamefont {Fan}}, \bibinfo {author} {\bibfnamefont
  {X.}~\bibnamefont {Hu}}, \bibinfo {author} {\bibfnamefont {Y.~W.}\
  \bibnamefont {Tseng}}, \bibinfo {author} {\bibfnamefont {Y.~H.}\ \bibnamefont
  {Lee}}, \bibinfo {author} {\bibfnamefont {V.}~\bibnamefont {Menon}}, \ and\
  \bibinfo {author} {\bibfnamefont {S.~R.}\ \bibnamefont {Forrest}},\ }\href
  {\doibase 10.1021/acs.nanolett.7b00695} {\bibfield  {journal} {\bibinfo
  {journal} {Nano~Lett.~}\ }\textbf {\bibinfo {volume} {17}},\ \bibinfo {pages}
  {3176} (\bibinfo {year} {2017})}\BibitemShut {NoStop}%
\bibitem [{\citenamefont {Huang}\ \emph {et~al.}(2018)\citenamefont {Huang},
  \citenamefont {Zhuge}, \citenamefont {Hou}, \citenamefont {Lv}, \citenamefont
  {Luo}, \citenamefont {Zhou}, \citenamefont {Gan},\ and\ \citenamefont
  {Zhai}}]{Huang2018}%
  \BibitemOpen
  \bibfield  {author} {\bibinfo {author} {\bibfnamefont {Y.}~\bibnamefont
  {Huang}}, \bibinfo {author} {\bibfnamefont {F.}~\bibnamefont {Zhuge}},
  \bibinfo {author} {\bibfnamefont {J.}~\bibnamefont {Hou}}, \bibinfo {author}
  {\bibfnamefont {L.}~\bibnamefont {Lv}}, \bibinfo {author} {\bibfnamefont
  {P.}~\bibnamefont {Luo}}, \bibinfo {author} {\bibfnamefont {N.}~\bibnamefont
  {Zhou}}, \bibinfo {author} {\bibfnamefont {L.}~\bibnamefont {Gan}}, \ and\
  \bibinfo {author} {\bibfnamefont {T.}~\bibnamefont {Zhai}},\ }\href {\doibase
  10.1021/acsnano.8b02380} {\bibfield  {journal} {\bibinfo  {journal}
  {ACS~Nano}\ }\textbf {\bibinfo {volume} {12}},\ \bibinfo {pages} {4062}
  (\bibinfo {year} {2018})}\BibitemShut {NoStop}%
\bibitem [{\citenamefont {Park}\ \emph {et~al.}(2021)\citenamefont {Park},
  \citenamefont {Mutz}, \citenamefont {Kovalenko}, \citenamefont {Schultz},
  \citenamefont {Shin}, \citenamefont {Aljarb}, \citenamefont {Li},
  \citenamefont {Tung}, \citenamefont {Amsalem}, \citenamefont
  {List-Kratochvil}, \citenamefont {St{\"{a}}hler}, \citenamefont {Xu},
  \citenamefont {Blumstengel},\ and\ \citenamefont {Koch}}]{Park2021}%
  \BibitemOpen
  \bibfield  {author} {\bibinfo {author} {\bibfnamefont {S.}~\bibnamefont
  {Park}}, \bibinfo {author} {\bibfnamefont {N.}~\bibnamefont {Mutz}}, \bibinfo
  {author} {\bibfnamefont {S.~A.}\ \bibnamefont {Kovalenko}}, \bibinfo {author}
  {\bibfnamefont {T.}~\bibnamefont {Schultz}}, \bibinfo {author} {\bibfnamefont
  {D.}~\bibnamefont {Shin}}, \bibinfo {author} {\bibfnamefont {A.}~\bibnamefont
  {Aljarb}}, \bibinfo {author} {\bibfnamefont {L.~J.}\ \bibnamefont {Li}},
  \bibinfo {author} {\bibfnamefont {V.}~\bibnamefont {Tung}}, \bibinfo {author}
  {\bibfnamefont {P.}~\bibnamefont {Amsalem}}, \bibinfo {author} {\bibfnamefont
  {E.~J.}\ \bibnamefont {List-Kratochvil}}, \bibinfo {author} {\bibfnamefont
  {J.}~\bibnamefont {St{\"{a}}hler}}, \bibinfo {author} {\bibfnamefont
  {X.}~\bibnamefont {Xu}}, \bibinfo {author} {\bibfnamefont {S.}~\bibnamefont
  {Blumstengel}}, \ and\ \bibinfo {author} {\bibfnamefont {N.}~\bibnamefont
  {Koch}},\ }\href {\doibase 10.1002/advs.202100215} {\bibfield  {journal}
  {\bibinfo  {journal} {Adv.~Sci.~}\ }\textbf {\bibinfo {volume} {8}},\
  \bibinfo {pages} {2100215} (\bibinfo {year} {2021})}\BibitemShut {NoStop}%
\bibitem [{\citenamefont {Jing}\ \emph {et~al.}(2014)\citenamefont {Jing},
  \citenamefont {Tan}, \citenamefont {Zhou},\ and\ \citenamefont
  {Shen}}]{Jing2014}%
  \BibitemOpen
  \bibfield  {author} {\bibinfo {author} {\bibfnamefont {Y.}~\bibnamefont
  {Jing}}, \bibinfo {author} {\bibfnamefont {X.}~\bibnamefont {Tan}}, \bibinfo
  {author} {\bibfnamefont {Z.}~\bibnamefont {Zhou}}, \ and\ \bibinfo {author}
  {\bibfnamefont {P.}~\bibnamefont {Shen}},\ }\href {\doibase
  10.1039/c4ta03660c} {\bibfield  {journal} {\bibinfo  {journal}
  {J.~Mater.~Chem.~A~}\ }\textbf {\bibinfo {volume} {2}},\ \bibinfo {pages}
  {16892} (\bibinfo {year} {2014})}\BibitemShut {NoStop}%
\bibitem [{\citenamefont {Wang}\ and\ \citenamefont {Paulus}(2020)}]{Wang2020}%
  \BibitemOpen
  \bibfield  {author} {\bibinfo {author} {\bibfnamefont {K.}~\bibnamefont
  {Wang}}\ and\ \bibinfo {author} {\bibfnamefont {B.}~\bibnamefont {Paulus}},\
  }\href {\doibase 10.1039/d0cp01239d} {\bibfield  {journal} {\bibinfo
  {journal} {Phys.~Chem.~Chem.~Phys.~}\ }\textbf {\bibinfo {volume} {22}},\
  \bibinfo {pages} {11936} (\bibinfo {year} {2020})}\BibitemShut {NoStop}%
\bibitem [{\citenamefont {Mutz}\ \emph {et~al.}(2020)\citenamefont {Mutz},
  \citenamefont {Park}, \citenamefont {Schultz}, \citenamefont {Sadofev},
  \citenamefont {Dalgleish}, \citenamefont {Reissig}, \citenamefont {Koch},
  \citenamefont {List-Kratochvil},\ and\ \citenamefont
  {Blumstengel}}]{Mutz2020}%
  \BibitemOpen
  \bibfield  {author} {\bibinfo {author} {\bibfnamefont {N.}~\bibnamefont
  {Mutz}}, \bibinfo {author} {\bibfnamefont {S.}~\bibnamefont {Park}}, \bibinfo
  {author} {\bibfnamefont {T.}~\bibnamefont {Schultz}}, \bibinfo {author}
  {\bibfnamefont {S.}~\bibnamefont {Sadofev}}, \bibinfo {author} {\bibfnamefont
  {S.}~\bibnamefont {Dalgleish}}, \bibinfo {author} {\bibfnamefont
  {L.}~\bibnamefont {Reissig}}, \bibinfo {author} {\bibfnamefont
  {N.}~\bibnamefont {Koch}}, \bibinfo {author} {\bibfnamefont {E.~J.}\
  \bibnamefont {List-Kratochvil}}, \ and\ \bibinfo {author} {\bibfnamefont
  {S.}~\bibnamefont {Blumstengel}},\ }\href {\doibase 10.1021/acs.jpcc.9b10877}
  {\bibfield  {journal} {\bibinfo  {journal} {J.~Phys.~Chem.~C}\ }\textbf
  {\bibinfo {volume} {124}},\ \bibinfo {pages} {2837} (\bibinfo {year}
  {2020})}\BibitemShut {NoStop}%
\bibitem [{\citenamefont {Canton-Vitoria}\ \emph {et~al.}(2020)\citenamefont
  {Canton-Vitoria}, \citenamefont {Sayed-Ahmad-baraza}, \citenamefont
  {Humbert}, \citenamefont {Arenal}, \citenamefont {Ewels},\ and\ \citenamefont
  {Tagmatarchis}}]{Canton-Vitoria2020}%
  \BibitemOpen
  \bibfield  {author} {\bibinfo {author} {\bibfnamefont {R.}~\bibnamefont
  {Canton-Vitoria}}, \bibinfo {author} {\bibfnamefont {Y.}~\bibnamefont
  {Sayed-Ahmad-baraza}}, \bibinfo {author} {\bibfnamefont {B.}~\bibnamefont
  {Humbert}}, \bibinfo {author} {\bibfnamefont {R.}~\bibnamefont {Arenal}},
  \bibinfo {author} {\bibfnamefont {C.~P.}\ \bibnamefont {Ewels}}, \ and\
  \bibinfo {author} {\bibfnamefont {N.}~\bibnamefont {Tagmatarchis}},\ }\href
  {\doibase 10.3390/nano10020363} {\bibfield  {journal} {\bibinfo  {journal}
  {Nanomaterials}\ }\textbf {\bibinfo {volume} {10}},\ \bibinfo {pages} {363}
  (\bibinfo {year} {2020})}\BibitemShut {NoStop}%
\bibitem [{\citenamefont {Yousofnejad}\ \emph {et~al.}(2020)\citenamefont
  {Yousofnejad}, \citenamefont {Reecht}, \citenamefont {Krane}, \citenamefont
  {Lotze},\ and\ \citenamefont {Franke}}]{Yousofnejad2020}%
  \BibitemOpen
  \bibfield  {author} {\bibinfo {author} {\bibfnamefont {A.}~\bibnamefont
  {Yousofnejad}}, \bibinfo {author} {\bibfnamefont {G.}~\bibnamefont {Reecht}},
  \bibinfo {author} {\bibfnamefont {N.}~\bibnamefont {Krane}}, \bibinfo
  {author} {\bibfnamefont {C.}~\bibnamefont {Lotze}}, \ and\ \bibinfo {author}
  {\bibfnamefont {K.~J.}\ \bibnamefont {Franke}},\ }\href {\doibase
  10.3762/BJNANO.11.91} {\bibfield  {journal} {\bibinfo  {journal} {Beilstein
  J. Nanotechnol.}\ }\textbf {\bibinfo {volume} {11}},\ \bibinfo {pages} {1062}
  (\bibinfo {year} {2020})}\BibitemShut {NoStop}%
\bibitem [{\citenamefont {Hybertsen}\ and\ \citenamefont
  {Louie}(1986)}]{Hybertsen1986}%
  \BibitemOpen
  \bibfield  {author} {\bibinfo {author} {\bibfnamefont {M.~S.}\ \bibnamefont
  {Hybertsen}}\ and\ \bibinfo {author} {\bibfnamefont {S.~G.}\ \bibnamefont
  {Louie}},\ }\href {\doibase 10.1103/PhysRevB.34.5390} {\bibfield  {journal}
  {\bibinfo  {journal} {Phys.~Rev.~B}\ }\textbf {\bibinfo {volume} {34}},\
  \bibinfo {pages} {5390} (\bibinfo {year} {1986})}\BibitemShut {NoStop}%
\bibitem [{\citenamefont {Rohlfing}\ and\ \citenamefont
  {Louie}(2000)}]{Rohlfing2000}%
  \BibitemOpen
  \bibfield  {author} {\bibinfo {author} {\bibfnamefont {M.}~\bibnamefont
  {Rohlfing}}\ and\ \bibinfo {author} {\bibfnamefont {S.~G.}\ \bibnamefont
  {Louie}},\ }\href
  {https://journals.aps.org/prb/abstract/10.1103/PhysRevB.62.4927} {\bibfield
  {journal} {\bibinfo  {journal} {Phys.~Rev.~B}\ }\textbf {\bibinfo {volume}
  {62}},\ \bibinfo {pages} {4927} (\bibinfo {year} {2000})}\BibitemShut
  {NoStop}%
\bibitem [{\citenamefont {Neaton}\ \emph {et~al.}(2006)\citenamefont {Neaton},
  \citenamefont {Hybertsen},\ and\ \citenamefont {Louie}}]{Neaton2006}%
  \BibitemOpen
  \bibfield  {author} {\bibinfo {author} {\bibfnamefont {J.~B.}\ \bibnamefont
  {Neaton}}, \bibinfo {author} {\bibfnamefont {M.~S.}\ \bibnamefont
  {Hybertsen}}, \ and\ \bibinfo {author} {\bibfnamefont {S.~G.}\ \bibnamefont
  {Louie}},\ }\href {\doibase 10.1103/PhysRevLett.97.216405} {\bibfield
  {journal} {\bibinfo  {journal} {Phys.~Rev.~Lett.~}\ }\textbf {\bibinfo
  {volume} {97}},\ \bibinfo {pages} {216405} (\bibinfo {year}
  {2006})}\BibitemShut {NoStop}%
\bibitem [{\citenamefont {Thygesen}\ and\ \citenamefont
  {Rubio}(2009)}]{Thygesen2009}%
  \BibitemOpen
  \bibfield  {author} {\bibinfo {author} {\bibfnamefont {K.~S.}\ \bibnamefont
  {Thygesen}}\ and\ \bibinfo {author} {\bibfnamefont {A.}~\bibnamefont
  {Rubio}},\ }\href {\doibase 10.1103/PhysRevLett.102.046802} {\bibfield
  {journal} {\bibinfo  {journal} {Phys.~Rev.~Lett.~}\ }\textbf {\bibinfo
  {volume} {102}},\ \bibinfo {pages} {046802} (\bibinfo {year}
  {2009})}\BibitemShut {NoStop}%
\bibitem [{\citenamefont {Egger}\ \emph {et~al.}(2015)\citenamefont {Egger},
  \citenamefont {Liu}, \citenamefont {Neaton},\ and\ \citenamefont
  {Kronik}}]{Egger2015}%
  \BibitemOpen
  \bibfield  {author} {\bibinfo {author} {\bibfnamefont {D.~A.}\ \bibnamefont
  {Egger}}, \bibinfo {author} {\bibfnamefont {Z.~F.}\ \bibnamefont {Liu}},
  \bibinfo {author} {\bibfnamefont {J.~B.}\ \bibnamefont {Neaton}}, \ and\
  \bibinfo {author} {\bibfnamefont {L.}~\bibnamefont {Kronik}},\ }\href
  {\doibase 10.1021/nl504863r} {\bibfield  {journal} {\bibinfo  {journal}
  {Nano~Lett.~}\ }\textbf {\bibinfo {volume} {15}},\ \bibinfo {pages} {2448}
  (\bibinfo {year} {2015})}\BibitemShut {NoStop}%
\bibitem [{\citenamefont {Fu}\ \emph {et~al.}(2016)\citenamefont {Fu},
  \citenamefont {Nabok},\ and\ \citenamefont {Draxl}}]{Fu2016}%
  \BibitemOpen
  \bibfield  {author} {\bibinfo {author} {\bibfnamefont {Q.}~\bibnamefont
  {Fu}}, \bibinfo {author} {\bibfnamefont {D.}~\bibnamefont {Nabok}}, \ and\
  \bibinfo {author} {\bibfnamefont {C.}~\bibnamefont {Draxl}},\ }\href
  {\doibase 10.1021/acs.jpcc.6b01741} {\bibfield  {journal} {\bibinfo
  {journal} {J.~Phys.~Chem.~C}\ }\textbf {\bibinfo {volume} {120}},\ \bibinfo
  {pages} {11671} (\bibinfo {year} {2016})}\BibitemShut {NoStop}%
\bibitem [{\citenamefont {Liu}\ \emph {et~al.}(2019)\citenamefont {Liu},
  \citenamefont {{Da Jornada}}, \citenamefont {Louie},\ and\ \citenamefont
  {Neaton}}]{Liu2019}%
  \BibitemOpen
  \bibfield  {author} {\bibinfo {author} {\bibfnamefont {Z.~F.}\ \bibnamefont
  {Liu}}, \bibinfo {author} {\bibfnamefont {F.~H.}\ \bibnamefont {{Da
  Jornada}}}, \bibinfo {author} {\bibfnamefont {S.~G.}\ \bibnamefont {Louie}},
  \ and\ \bibinfo {author} {\bibfnamefont {J.~B.}\ \bibnamefont {Neaton}},\
  }\href {\doibase 10.1021/acs.jctc.9b00326} {\bibfield  {journal} {\bibinfo
  {journal} {J.~Chem.~Theory.~Comput.~}\ }\textbf {\bibinfo {volume} {15}},\
  \bibinfo {pages} {4218} (\bibinfo {year} {2019})}\BibitemShut {NoStop}%
\bibitem [{\citenamefont {Nabok}\ \emph {et~al.}(2019)\citenamefont {Nabok},
  \citenamefont {H{\"{o}}ffling},\ and\ \citenamefont {Draxl}}]{Nabok2019}%
  \BibitemOpen
  \bibfield  {author} {\bibinfo {author} {\bibfnamefont {D.}~\bibnamefont
  {Nabok}}, \bibinfo {author} {\bibfnamefont {B.}~\bibnamefont
  {H{\"{o}}ffling}}, \ and\ \bibinfo {author} {\bibfnamefont {C.}~\bibnamefont
  {Draxl}},\ }\href {\doibase 10.1021/acs.chemmater.9b01802} {\bibfield
  {journal} {\bibinfo  {journal} {Chem.~Mater.~}\ }\textbf {\bibinfo {volume}
  {31}},\ \bibinfo {pages} {7143} (\bibinfo {year} {2019})}\BibitemShut
  {NoStop}%
\bibitem [{\citenamefont {Bernardi}\ \emph {et~al.}(2013)\citenamefont
  {Bernardi}, \citenamefont {Palummo},\ and\ \citenamefont
  {Grossman}}]{Bernardi2013}%
  \BibitemOpen
  \bibfield  {author} {\bibinfo {author} {\bibfnamefont {M.}~\bibnamefont
  {Bernardi}}, \bibinfo {author} {\bibfnamefont {M.}~\bibnamefont {Palummo}}, \
  and\ \bibinfo {author} {\bibfnamefont {J.~C.}\ \bibnamefont {Grossman}},\
  }\href {\doibase 10.1021/nl401544y} {\bibfield  {journal} {\bibinfo
  {journal} {Nano~Lett.~}\ }\textbf {\bibinfo {volume} {13}},\ \bibinfo {pages}
  {3664} (\bibinfo {year} {2013})}\BibitemShut {NoStop}%
\bibitem [{\citenamefont {Turkina}\ \emph {et~al.}(2019)\citenamefont
  {Turkina}, \citenamefont {Nabok}, \citenamefont {Gulans}, \citenamefont
  {Cocchi},\ and\ \citenamefont {Draxl}}]{Turkina2018}%
  \BibitemOpen
  \bibfield  {author} {\bibinfo {author} {\bibfnamefont {O.}~\bibnamefont
  {Turkina}}, \bibinfo {author} {\bibfnamefont {D.}~\bibnamefont {Nabok}},
  \bibinfo {author} {\bibfnamefont {A.}~\bibnamefont {Gulans}}, \bibinfo
  {author} {\bibfnamefont {C.}~\bibnamefont {Cocchi}}, \ and\ \bibinfo {author}
  {\bibfnamefont {C.}~\bibnamefont {Draxl}},\ }\href {\doibase
  10.1002/adts.201800108} {\bibfield  {journal} {\bibinfo  {journal}
  {Adv.~Theory~Simul.~}\ }\textbf {\bibinfo {volume} {2}},\ \bibinfo {pages}
  {1800108} (\bibinfo {year} {2019})}\BibitemShut {NoStop}%
\bibitem [{\citenamefont {Aggoune}\ \emph {et~al.}(2020)\citenamefont
  {Aggoune}, \citenamefont {Cocchi}, \citenamefont {Nabok}, \citenamefont
  {Rezouali}, \citenamefont {Belkhir},\ and\ \citenamefont
  {Draxl}}]{Aggoune2020}%
  \BibitemOpen
  \bibfield  {author} {\bibinfo {author} {\bibfnamefont {W.}~\bibnamefont
  {Aggoune}}, \bibinfo {author} {\bibfnamefont {C.}~\bibnamefont {Cocchi}},
  \bibinfo {author} {\bibfnamefont {D.}~\bibnamefont {Nabok}}, \bibinfo
  {author} {\bibfnamefont {K.}~\bibnamefont {Rezouali}}, \bibinfo {author}
  {\bibfnamefont {M.~A.}\ \bibnamefont {Belkhir}}, \ and\ \bibinfo {author}
  {\bibfnamefont {C.}~\bibnamefont {Draxl}},\ }\href {\doibase
  10.1103/physrevmaterials.4.084001} {\bibfield  {journal} {\bibinfo  {journal}
  {Phys.~Rev.~Mater.~}\ }\textbf {\bibinfo {volume} {4}},\ \bibinfo {pages}
  {84001} (\bibinfo {year} {2020})}\BibitemShut {NoStop}%
\bibitem [{\citenamefont {Edalati-Boostan}\ \emph {et~al.}(2020)\citenamefont
  {Edalati-Boostan}, \citenamefont {Cocchi},\ and\ \citenamefont
  {Draxl}}]{Edalati-Boostan2020}%
  \BibitemOpen
  \bibfield  {author} {\bibinfo {author} {\bibfnamefont {S.}~\bibnamefont
  {Edalati-Boostan}}, \bibinfo {author} {\bibfnamefont {C.}~\bibnamefont
  {Cocchi}}, \ and\ \bibinfo {author} {\bibfnamefont {C.}~\bibnamefont
  {Draxl}},\ }\href {\doibase 10.1103/PhysRevMaterials.4.085202} {\bibfield
  {journal} {\bibinfo  {journal} {Phys.~Rev.~Mater.~}\ }\textbf {\bibinfo
  {volume} {4}},\ \bibinfo {pages} {085202} (\bibinfo {year}
  {2020})}\BibitemShut {NoStop}%
\bibitem [{\citenamefont {Kang}\ \emph {et~al.}(2018)\citenamefont {Kang},
  \citenamefont {Jung}, \citenamefont {Shin}, \citenamefont {Sohn},
  \citenamefont {Ryu}, \citenamefont {Kim}, \citenamefont {Hoesch},\ and\
  \citenamefont {Kim}}]{Kang2018}%
  \BibitemOpen
  \bibfield  {author} {\bibinfo {author} {\bibfnamefont {M.}~\bibnamefont
  {Kang}}, \bibinfo {author} {\bibfnamefont {S.~W.}\ \bibnamefont {Jung}},
  \bibinfo {author} {\bibfnamefont {W.~J.}\ \bibnamefont {Shin}}, \bibinfo
  {author} {\bibfnamefont {Y.}~\bibnamefont {Sohn}}, \bibinfo {author}
  {\bibfnamefont {S.~H.}\ \bibnamefont {Ryu}}, \bibinfo {author} {\bibfnamefont
  {T.~K.}\ \bibnamefont {Kim}}, \bibinfo {author} {\bibfnamefont
  {M.}~\bibnamefont {Hoesch}}, \ and\ \bibinfo {author} {\bibfnamefont {K.~S.}\
  \bibnamefont {Kim}},\ }\href {\doibase 10.1038/s41563-018-0092-7} {\bibfield
  {journal} {\bibinfo  {journal} {Nat.~Mater.}\ }\textbf {\bibinfo {volume}
  {17}},\ \bibinfo {pages} {676} (\bibinfo {year} {2018})}\BibitemShut
  {NoStop}%
\bibitem [{\citenamefont {Caruso}\ \emph {et~al.}(2021)\citenamefont {Caruso},
  \citenamefont {Amsalem}, \citenamefont {Ma}, \citenamefont {Aljarb},
  \citenamefont {Schultz}, \citenamefont {Zacharias}, \citenamefont {Tung},
  \citenamefont {Koch},\ and\ \citenamefont {Draxl}}]{Caruso2021B}%
  \BibitemOpen
  \bibfield  {author} {\bibinfo {author} {\bibfnamefont {F.}~\bibnamefont
  {Caruso}}, \bibinfo {author} {\bibfnamefont {P.}~\bibnamefont {Amsalem}},
  \bibinfo {author} {\bibfnamefont {J.}~\bibnamefont {Ma}}, \bibinfo {author}
  {\bibfnamefont {A.}~\bibnamefont {Aljarb}}, \bibinfo {author} {\bibfnamefont
  {T.}~\bibnamefont {Schultz}}, \bibinfo {author} {\bibfnamefont
  {M.}~\bibnamefont {Zacharias}}, \bibinfo {author} {\bibfnamefont
  {V.}~\bibnamefont {Tung}}, \bibinfo {author} {\bibfnamefont {N.}~\bibnamefont
  {Koch}}, \ and\ \bibinfo {author} {\bibfnamefont {C.}~\bibnamefont {Draxl}},\
  }\href {\doibase 10.1103/PhysRevB.103.205152} {\bibfield  {journal} {\bibinfo
   {journal} {Phys.~Rev.~B}\ }\textbf {\bibinfo {volume} {103}},\ \bibinfo
  {pages} {205152} (\bibinfo {year} {2021})}\BibitemShut {NoStop}%
\bibitem [{\citenamefont {Pisarra}\ \emph {et~al.}(2021)\citenamefont
  {Pisarra}, \citenamefont {Di{\'{a}}z},\ and\ \citenamefont
  {Mart{\'{i}}n}}]{Pisarra2021}%
  \BibitemOpen
  \bibfield  {author} {\bibinfo {author} {\bibfnamefont {M.}~\bibnamefont
  {Pisarra}}, \bibinfo {author} {\bibfnamefont {C.}~\bibnamefont {Di{\'{a}}z}},
  \ and\ \bibinfo {author} {\bibfnamefont {F.}~\bibnamefont {Mart{\'{i}}n}},\
  }\href {\doibase 10.1103/PhysRevB.103.195416} {\bibfield  {journal} {\bibinfo
   {journal} {Phys.~Rev.~B}\ }\textbf {\bibinfo {volume} {103}},\ \bibinfo
  {pages} {195416} (\bibinfo {year} {2021})}\BibitemShut {NoStop}%
\bibitem [{\citenamefont {Hohenberg}\ and\ \citenamefont
  {Kohn}(1964)}]{Hohenberg1964}%
  \BibitemOpen
  \bibfield  {author} {\bibinfo {author} {\bibfnamefont {P.}~\bibnamefont
  {Hohenberg}}\ and\ \bibinfo {author} {\bibfnamefont {W.}~\bibnamefont
  {Kohn}},\ }\href {\doibase 10.1103/PhysRevB.7.1912} {\bibfield  {journal}
  {\bibinfo  {journal} {Phys.~Rev.~}\ }\textbf {\bibinfo {volume} {136}},\
  \bibinfo {pages} {B864} (\bibinfo {year} {1964})}\BibitemShut {NoStop}%
\bibitem [{\citenamefont {Kohn}\ and\ \citenamefont {Sham}(1965)}]{Kohn1965}%
  \BibitemOpen
  \bibfield  {author} {\bibinfo {author} {\bibfnamefont {W.}~\bibnamefont
  {Kohn}}\ and\ \bibinfo {author} {\bibfnamefont {L.~J.}\ \bibnamefont
  {Sham}},\ }\href {\doibase 10.1103/PhysRev.140.A1133} {\bibfield  {journal}
  {\bibinfo  {journal} {Phys.~Rev.~}\ }\textbf {\bibinfo {volume} {140}},\
  \bibinfo {pages} {A1133} (\bibinfo {year} {1965})}\BibitemShut {NoStop}%
\bibitem [{\citenamefont {Perdew}\ \emph {et~al.}(1996)\citenamefont {Perdew},
  \citenamefont {Burke},\ and\ \citenamefont {Ernzerhof}}]{Perdew1996}%
  \BibitemOpen
  \bibfield  {author} {\bibinfo {author} {\bibfnamefont {J.~P.}\ \bibnamefont
  {Perdew}}, \bibinfo {author} {\bibfnamefont {K.}~\bibnamefont {Burke}}, \
  and\ \bibinfo {author} {\bibfnamefont {M.}~\bibnamefont {Ernzerhof}},\ }\href
  {\doibase 10.1103/PhysRevLett.77.3865} {\bibfield  {journal} {\bibinfo
  {journal} {Phys.~Rev.~Lett.~}\ }\textbf {\bibinfo {volume} {77}},\ \bibinfo
  {pages} {3865} (\bibinfo {year} {1996})}\BibitemShut {NoStop}%
\bibitem [{\citenamefont {Tkatchenko}\ and\ \citenamefont
  {Scheffler}(2009)}]{Tkatchenko2009}%
  \BibitemOpen
  \bibfield  {author} {\bibinfo {author} {\bibfnamefont {A.}~\bibnamefont
  {Tkatchenko}}\ and\ \bibinfo {author} {\bibfnamefont {M.}~\bibnamefont
  {Scheffler}},\ }\href {\doibase 10.1103/PhysRevLett.102.073005} {\bibfield
  {journal} {\bibinfo  {journal} {Phys.~Rev.~Lett.~}\ }\textbf {\bibinfo
  {volume} {102}},\ \bibinfo {pages} {073005} (\bibinfo {year}
  {2009})}\BibitemShut {NoStop}%
\bibitem [{\citenamefont {Blum}\ \emph {et~al.}(2009)\citenamefont {Blum},
  \citenamefont {Gehrke}, \citenamefont {Hanke}, \citenamefont {Havu},
  \citenamefont {Havu}, \citenamefont {Ren}, \citenamefont {Reuter},\ and\
  \citenamefont {Scheffler}}]{Blum2009}%
  \BibitemOpen
  \bibfield  {author} {\bibinfo {author} {\bibfnamefont {V.}~\bibnamefont
  {Blum}}, \bibinfo {author} {\bibfnamefont {R.}~\bibnamefont {Gehrke}},
  \bibinfo {author} {\bibfnamefont {F.}~\bibnamefont {Hanke}}, \bibinfo
  {author} {\bibfnamefont {P.}~\bibnamefont {Havu}}, \bibinfo {author}
  {\bibfnamefont {V.}~\bibnamefont {Havu}}, \bibinfo {author} {\bibfnamefont
  {X.}~\bibnamefont {Ren}}, \bibinfo {author} {\bibfnamefont {K.}~\bibnamefont
  {Reuter}}, \ and\ \bibinfo {author} {\bibfnamefont {M.}~\bibnamefont
  {Scheffler}},\ }\href {\doibase 10.1016/j.cpc.2009.06.022} {\bibfield
  {journal} {\bibinfo  {journal} {Comput.~Phys.~Commun.~}\ }\textbf {\bibinfo
  {volume} {180}},\ \bibinfo {pages} {2175} (\bibinfo {year}
  {2009})}\BibitemShut {NoStop}%
\bibitem [{\citenamefont {Gulans}\ \emph {et~al.}(2014)\citenamefont {Gulans},
  \citenamefont {Kontur}, \citenamefont {Meisenbichler}, \citenamefont {Nabok},
  \citenamefont {Pavone}, \citenamefont {Rigamonti}, \citenamefont
  {Sagmeister}, \citenamefont {Werner},\ and\ \citenamefont
  {Draxl}}]{Gulans2014}%
  \BibitemOpen
  \bibfield  {author} {\bibinfo {author} {\bibfnamefont {A.}~\bibnamefont
  {Gulans}}, \bibinfo {author} {\bibfnamefont {S.}~\bibnamefont {Kontur}},
  \bibinfo {author} {\bibfnamefont {C.}~\bibnamefont {Meisenbichler}}, \bibinfo
  {author} {\bibfnamefont {D.}~\bibnamefont {Nabok}}, \bibinfo {author}
  {\bibfnamefont {P.}~\bibnamefont {Pavone}}, \bibinfo {author} {\bibfnamefont
  {S.}~\bibnamefont {Rigamonti}}, \bibinfo {author} {\bibfnamefont
  {S.}~\bibnamefont {Sagmeister}}, \bibinfo {author} {\bibfnamefont
  {U.}~\bibnamefont {Werner}}, \ and\ \bibinfo {author} {\bibfnamefont
  {C.}~\bibnamefont {Draxl}},\ }\href {\doibase 10.1088/0953-8984/26/36/363202}
  {\bibfield  {journal} {\bibinfo  {journal} {J.~Phys.~Condens.~Matter.~}\
  }\textbf {\bibinfo {volume} {26}},\ \bibinfo {pages} {363202} (\bibinfo
  {year} {2014})}\BibitemShut {NoStop}%
\bibitem [{\citenamefont {Nabok}\ \emph {et~al.}(2016)\citenamefont {Nabok},
  \citenamefont {Gulans},\ and\ \citenamefont {Draxl}}]{Nabok2016}%
  \BibitemOpen
  \bibfield  {author} {\bibinfo {author} {\bibfnamefont {D.}~\bibnamefont
  {Nabok}}, \bibinfo {author} {\bibfnamefont {A.}~\bibnamefont {Gulans}}, \
  and\ \bibinfo {author} {\bibfnamefont {C.}~\bibnamefont {Draxl}},\ }\href
  {\doibase 10.1103/PhysRevB.94.035118} {\bibfield  {journal} {\bibinfo
  {journal} {Phys.~Rev.~B}\ }\textbf {\bibinfo {volume} {94}},\ \bibinfo
  {pages} {035118} (\bibinfo {year} {2016})}\BibitemShut {NoStop}%
\bibitem [{\citenamefont {Vorwerk}\ \emph {et~al.}(2019)\citenamefont
  {Vorwerk}, \citenamefont {Aurich}, \citenamefont {Cocchi},\ and\
  \citenamefont {Draxl}}]{Vorwerk2019}%
  \BibitemOpen
  \bibfield  {author} {\bibinfo {author} {\bibfnamefont {C.}~\bibnamefont
  {Vorwerk}}, \bibinfo {author} {\bibfnamefont {B.}~\bibnamefont {Aurich}},
  \bibinfo {author} {\bibfnamefont {C.}~\bibnamefont {Cocchi}}, \ and\ \bibinfo
  {author} {\bibfnamefont {C.}~\bibnamefont {Draxl}},\ }\href {\doibase
  10.1088/2516-1075/ab3123} {\bibfield  {journal} {\bibinfo  {journal}
  {Electron.~Struct.~}\ }\textbf {\bibinfo {volume} {1}},\ \bibinfo {pages}
  {037001} (\bibinfo {year} {2019})}\BibitemShut {NoStop}%
\bibitem [{\citenamefont {Hedin}(1965)}]{Hedin1965}%
  \BibitemOpen
  \bibfield  {author} {\bibinfo {author} {\bibfnamefont {L.}~\bibnamefont
  {Hedin}},\ }\href {\doibase 10.1103/PhysRev.139.A796} {\bibfield  {journal}
  {\bibinfo  {journal} {Phys.~Rev.~}\ }\textbf {\bibinfo {volume} {139}},\
  \bibinfo {pages} {A796} (\bibinfo {year} {1965})}\BibitemShut {NoStop}%
\bibitem [{\citenamefont {Hybertsen}\ and\ \citenamefont
  {Louie}(1985)}]{Hybertsen1985}%
  \BibitemOpen
  \bibfield  {author} {\bibinfo {author} {\bibfnamefont {M.~S.}\ \bibnamefont
  {Hybertsen}}\ and\ \bibinfo {author} {\bibfnamefont {S.~G.}\ \bibnamefont
  {Louie}},\ }\href {\doibase 10.1103/PhysRevLett.55.1418} {\bibfield
  {journal} {\bibinfo  {journal} {Phys.~Rev.~Lett.~}\ }\textbf {\bibinfo
  {volume} {55}},\ \bibinfo {pages} {1418} (\bibinfo {year}
  {1985})}\BibitemShut {NoStop}%
\bibitem [{\citenamefont {Tillack}\ \emph {et~al.}(2020)\citenamefont
  {Tillack}, \citenamefont {Gulans},\ and\ \citenamefont
  {Draxl}}]{Tillack2020}%
  \BibitemOpen
  \bibfield  {author} {\bibinfo {author} {\bibfnamefont {S.}~\bibnamefont
  {Tillack}}, \bibinfo {author} {\bibfnamefont {A.}~\bibnamefont {Gulans}}, \
  and\ \bibinfo {author} {\bibfnamefont {C.}~\bibnamefont {Draxl}},\ }\href
  {\doibase 10.1103/PhysRevB.101.235102} {\bibfield  {journal} {\bibinfo
  {journal} {Phys.~Rev.~B}\ }\textbf {\bibinfo {volume} {101}},\ \bibinfo
  {pages} {235102} (\bibinfo {year} {2020})}\BibitemShut {NoStop}%
\bibitem [{\citenamefont {Hanke}\ and\ \citenamefont {Sham}(1980)}]{Hanke1980}%
  \BibitemOpen
  \bibfield  {author} {\bibinfo {author} {\bibfnamefont {W.}~\bibnamefont
  {Hanke}}\ and\ \bibinfo {author} {\bibfnamefont {L.~J.}\ \bibnamefont
  {Sham}},\ }\href {\doibase 10.1103/PhysRevB.21.4656} {\bibfield  {journal}
  {\bibinfo  {journal} {Phys.~Rev.~B}\ }\textbf {\bibinfo {volume} {21}},\
  \bibinfo {pages} {4656} (\bibinfo {year} {1980})}\BibitemShut {NoStop}%
\bibitem [{\citenamefont {Strinati}(1988)}]{Strinati1988}%
  \BibitemOpen
  \bibfield  {author} {\bibinfo {author} {\bibfnamefont {G.}~\bibnamefont
  {Strinati}},\ }\href {\doibase 10.1007/BF02725962} {\bibfield  {journal}
  {\bibinfo  {journal} {La Rivista Del Nuovo Cimento Series 3}\ }\textbf
  {\bibinfo {volume} {11}},\ \bibinfo {pages} {1} (\bibinfo {year}
  {1988})}\BibitemShut {NoStop}%
\bibitem [{\citenamefont {Momma}\ and\ \citenamefont
  {Izumi}(2011)}]{Momma2011}%
  \BibitemOpen
  \bibfield  {author} {\bibinfo {author} {\bibfnamefont {K.}~\bibnamefont
  {Momma}}\ and\ \bibinfo {author} {\bibfnamefont {F.}~\bibnamefont {Izumi}},\
  }\href {\doibase 10.1107/S0021889811038970} {\bibfield  {journal} {\bibinfo
  {journal} {J.~Appl.~Cryst.~}\ }\textbf {\bibinfo {volume} {44}},\ \bibinfo
  {pages} {1272} (\bibinfo {year} {2011})}\BibitemShut {NoStop}%
\bibitem [{\citenamefont {Draxl}\ and\ \citenamefont
  {Scheffler}(2019)}]{Draxl2019}%
  \BibitemOpen
  \bibfield  {author} {\bibinfo {author} {\bibfnamefont {C.}~\bibnamefont
  {Draxl}}\ and\ \bibinfo {author} {\bibfnamefont {M.}~\bibnamefont
  {Scheffler}},\ }\href {\doibase 10.1088/2515-7639/ab13bb} {\bibfield
  {journal} {\bibinfo  {journal} {J.~Phys.~Mater.}\ }\textbf {\bibinfo {volume}
  {2}},\ \bibinfo {pages} {036001} (\bibinfo {year} {2019})}\BibitemShut
  {NoStop}%
\bibitem [{\citenamefont {Gonz{\'{a}}lez}\ \emph {et~al.}(2017)\citenamefont
  {Gonz{\'{a}}lez}, \citenamefont {Biel},\ and\ \citenamefont
  {Dappe}}]{Gonzalez2017}%
  \BibitemOpen
  \bibfield  {author} {\bibinfo {author} {\bibfnamefont {C.}~\bibnamefont
  {Gonz{\'{a}}lez}}, \bibinfo {author} {\bibfnamefont {B.}~\bibnamefont
  {Biel}}, \ and\ \bibinfo {author} {\bibfnamefont {Y.~J.}\ \bibnamefont
  {Dappe}},\ }\href {\doibase 10.1039/c7cp00544j} {\bibfield  {journal}
  {\bibinfo  {journal} {Phys.~Chem.~Chem.~Phys.~}\ }\textbf {\bibinfo {volume}
  {19}},\ \bibinfo {pages} {9485} (\bibinfo {year} {2017})}\BibitemShut
  {NoStop}%
\bibitem [{\citenamefont {Valencia}\ and\ \citenamefont
  {Caldas}(2017)}]{Valencia2017}%
  \BibitemOpen
  \bibfield  {author} {\bibinfo {author} {\bibfnamefont {A.~M.}\ \bibnamefont
  {Valencia}}\ and\ \bibinfo {author} {\bibfnamefont {M.~J.}\ \bibnamefont
  {Caldas}},\ }\href {\doibase 10.1103/PhysRevB.96.125431} {\bibfield
  {journal} {\bibinfo  {journal} {Phys.~Rev.~B}\ }\textbf {\bibinfo {volume}
  {96}},\ \bibinfo {pages} {125431} (\bibinfo {year} {2017})}\BibitemShut
  {NoStop}%
\bibitem [{\citenamefont {Shirai}\ and\ \citenamefont
  {Inagaki}(2020)}]{Shirai2020}%
  \BibitemOpen
  \bibfield  {author} {\bibinfo {author} {\bibfnamefont {S.}~\bibnamefont
  {Shirai}}\ and\ \bibinfo {author} {\bibfnamefont {S.}~\bibnamefont
  {Inagaki}},\ }\href {\doibase 10.1039/c9ra10483f} {\bibfield  {journal}
  {\bibinfo  {journal} {RSC Advances}\ }\textbf {\bibinfo {volume} {10}},\
  \bibinfo {pages} {12988} (\bibinfo {year} {2020})}\BibitemShut {NoStop}%
\bibitem [{\citenamefont {Marom}\ \emph {et~al.}(2012)\citenamefont {Marom},
  \citenamefont {Caruso}, \citenamefont {Ren}, \citenamefont {Hofmann},
  \citenamefont {K{\"{o}}rzd{\"{o}}rfer}, \citenamefont {Chelikowsky},
  \citenamefont {Rubio}, \citenamefont {Scheffler},\ and\ \citenamefont
  {Rinke}}]{Marom2012}%
  \BibitemOpen
  \bibfield  {author} {\bibinfo {author} {\bibfnamefont {N.}~\bibnamefont
  {Marom}}, \bibinfo {author} {\bibfnamefont {F.}~\bibnamefont {Caruso}},
  \bibinfo {author} {\bibfnamefont {X.}~\bibnamefont {Ren}}, \bibinfo {author}
  {\bibfnamefont {O.~T.}\ \bibnamefont {Hofmann}}, \bibinfo {author}
  {\bibfnamefont {T.}~\bibnamefont {K{\"{o}}rzd{\"{o}}rfer}}, \bibinfo {author}
  {\bibfnamefont {J.~R.}\ \bibnamefont {Chelikowsky}}, \bibinfo {author}
  {\bibfnamefont {A.}~\bibnamefont {Rubio}}, \bibinfo {author} {\bibfnamefont
  {M.}~\bibnamefont {Scheffler}}, \ and\ \bibinfo {author} {\bibfnamefont
  {P.}~\bibnamefont {Rinke}},\ }\href {\doibase 10.1103/PhysRevB.86.245127}
  {\bibfield  {journal} {\bibinfo  {journal} {Phys.~Rev.~B}\ }\textbf {\bibinfo
  {volume} {86}},\ \bibinfo {pages} {245127} (\bibinfo {year}
  {2012})}\BibitemShut {NoStop}%
\bibitem [{\citenamefont {Jones}\ and\ \citenamefont
  {Gunnarsson}(1989)}]{Jones1989}%
  \BibitemOpen
  \bibfield  {author} {\bibinfo {author} {\bibfnamefont {R.}~\bibnamefont
  {Jones}}\ and\ \bibinfo {author} {\bibfnamefont {O.}~\bibnamefont
  {Gunnarsson}},\ }\href {\doibase 10.1103/RevModPhys.61.689} {\bibfield
  {journal} {\bibinfo  {journal} {Rev.~Mod.~Phys.~}\ }\textbf {\bibinfo
  {volume} {61}},\ \bibinfo {pages} {689 } (\bibinfo {year}
  {1989})}\BibitemShut {NoStop}%
\bibitem [{\citenamefont {Onida}\ \emph {et~al.}(2002)\citenamefont {Onida},
  \citenamefont {Reining},\ and\ \citenamefont {Rubio}}]{Onida2002}%
  \BibitemOpen
  \bibfield  {author} {\bibinfo {author} {\bibfnamefont {G.}~\bibnamefont
  {Onida}}, \bibinfo {author} {\bibfnamefont {L.}~\bibnamefont {Reining}}, \
  and\ \bibinfo {author} {\bibfnamefont {A.}~\bibnamefont {Rubio}},\ }\href
  {\doibase 10.1103/RevModPhys.74.601} {\bibfield  {journal} {\bibinfo
  {journal} {Rev.~Mod.~Phys.~}\ }\textbf {\bibinfo {volume} {74}},\ \bibinfo
  {pages} {601 } (\bibinfo {year} {2002})}\BibitemShut {NoStop}%
\bibitem [{\citenamefont {Cohen}\ \emph {et~al.}(2012)\citenamefont {Cohen},
  \citenamefont {Mori-S{\'{a}}nchez},\ and\ \citenamefont {Yang}}]{Cohen2012}%
  \BibitemOpen
  \bibfield  {author} {\bibinfo {author} {\bibfnamefont {A.~J.}\ \bibnamefont
  {Cohen}}, \bibinfo {author} {\bibfnamefont {P.}~\bibnamefont
  {Mori-S{\'{a}}nchez}}, \ and\ \bibinfo {author} {\bibfnamefont
  {W.}~\bibnamefont {Yang}},\ }\href {\doibase 10.1021/cr200107z} {\bibfield
  {journal} {\bibinfo  {journal} {Chem.~Rev.~}\ }\textbf {\bibinfo {volume}
  {112}},\ \bibinfo {pages} {289} (\bibinfo {year} {2012})}\BibitemShut
  {NoStop}%
\bibitem [{\citenamefont {Draxl}\ \emph {et~al.}(2014)\citenamefont {Draxl},
  \citenamefont {Nabok},\ and\ \citenamefont {Hannewald}}]{Draxl2014}%
  \BibitemOpen
  \bibfield  {author} {\bibinfo {author} {\bibfnamefont {C.}~\bibnamefont
  {Draxl}}, \bibinfo {author} {\bibfnamefont {D.}~\bibnamefont {Nabok}}, \ and\
  \bibinfo {author} {\bibfnamefont {K.}~\bibnamefont {Hannewald}},\ }\href
  {\doibase 10.1021/ar500096q} {\bibfield  {journal} {\bibinfo  {journal}
  {Acc.~Chem.~Res.~}\ }\textbf {\bibinfo {volume} {47}},\ \bibinfo {pages}
  {3225} (\bibinfo {year} {2014})}\BibitemShut {NoStop}%
\bibitem [{\citenamefont {Marom}(2017)}]{Marom2017}%
  \BibitemOpen
  \bibfield  {author} {\bibinfo {author} {\bibfnamefont {N.}~\bibnamefont
  {Marom}},\ }\href {\doibase 10.1088/1361-648X/29/10/103003} {\bibfield
  {journal} {\bibinfo  {journal} {J.~Phys.~Condens.~Matter.~}\ }\textbf
  {\bibinfo {volume} {29}},\ \bibinfo {pages} {103003} (\bibinfo {year}
  {2017})}\BibitemShut {NoStop}%
\bibitem [{\citenamefont {Caruso}\ \emph {et~al.}(2016)\citenamefont {Caruso},
  \citenamefont {Dauth}, \citenamefont {{Van Setten}},\ and\ \citenamefont
  {Rinke}}]{Caruso2016}%
  \BibitemOpen
  \bibfield  {author} {\bibinfo {author} {\bibfnamefont {F.}~\bibnamefont
  {Caruso}}, \bibinfo {author} {\bibfnamefont {M.}~\bibnamefont {Dauth}},
  \bibinfo {author} {\bibfnamefont {M.~J.}\ \bibnamefont {{Van Setten}}}, \
  and\ \bibinfo {author} {\bibfnamefont {P.}~\bibnamefont {Rinke}},\ }\href
  {\doibase 10.1021/acs.jctc.6b00774} {\bibfield  {journal} {\bibinfo
  {journal} {J.~Chem.~Theory.~Comput.~}\ }\textbf {\bibinfo {volume} {12}},\
  \bibinfo {pages} {5076} (\bibinfo {year} {2016})}\BibitemShut {NoStop}%
\bibitem [{\citenamefont {Dauth}\ \emph {et~al.}(2016)\citenamefont {Dauth},
  \citenamefont {Caruso}, \citenamefont {K{\"{u}}mmel},\ and\ \citenamefont
  {Rinke}}]{Dauth2016}%
  \BibitemOpen
  \bibfield  {author} {\bibinfo {author} {\bibfnamefont {M.}~\bibnamefont
  {Dauth}}, \bibinfo {author} {\bibfnamefont {F.}~\bibnamefont {Caruso}},
  \bibinfo {author} {\bibfnamefont {S.}~\bibnamefont {K{\"{u}}mmel}}, \ and\
  \bibinfo {author} {\bibfnamefont {P.}~\bibnamefont {Rinke}},\ }\href
  {\doibase 10.1103/PhysRevB.93.121115} {\bibfield  {journal} {\bibinfo
  {journal} {Phys.~Rev.~B}\ }\textbf {\bibinfo {volume} {93}},\ \bibinfo
  {pages} {121115(R)} (\bibinfo {year} {2016})}\BibitemShut {NoStop}%
\bibitem [{\citenamefont {Marsili}\ \emph {et~al.}(2021)\citenamefont
  {Marsili}, \citenamefont {Molina-S{\'{a}}nchez}, \citenamefont {Palummo},
  \citenamefont {Sangalli},\ and\ \citenamefont {Marini}}]{Marsili2021}%
  \BibitemOpen
  \bibfield  {author} {\bibinfo {author} {\bibfnamefont {M.}~\bibnamefont
  {Marsili}}, \bibinfo {author} {\bibfnamefont {A.}~\bibnamefont
  {Molina-S{\'{a}}nchez}}, \bibinfo {author} {\bibfnamefont {M.}~\bibnamefont
  {Palummo}}, \bibinfo {author} {\bibfnamefont {D.}~\bibnamefont {Sangalli}}, \
  and\ \bibinfo {author} {\bibfnamefont {A.}~\bibnamefont {Marini}},\ }\href
  {\doibase 10.1103/PhysRevB.103.155152} {\bibfield  {journal} {\bibinfo
  {journal} {Phys.~Rev.~B}\ }\textbf {\bibinfo {volume} {103}},\ \bibinfo
  {pages} {155152} (\bibinfo {year} {2021})}\BibitemShut {NoStop}%
\bibitem [{\citenamefont {Vona}(2021)}]{Vona2021}%
  \BibitemOpen
  \bibfield  {author} {\bibinfo {author} {\bibfnamefont {C.}~\bibnamefont
  {Vona}},\ }\href@noop {} {}\bibinfo {howpublished} {private communication}
  (\bibinfo {year} {2021})\BibitemShut {NoStop}%
\bibitem [{\citenamefont {Aggoune}\ \emph {et~al.}(2018)\citenamefont
  {Aggoune}, \citenamefont {Cocchi}, \citenamefont {Nabok}, \citenamefont
  {Rezouali}, \citenamefont {Belkhir},\ and\ \citenamefont
  {Draxl}}]{aggoune2018PRB}%
  \BibitemOpen
  \bibfield  {author} {\bibinfo {author} {\bibfnamefont {W.}~\bibnamefont
  {Aggoune}}, \bibinfo {author} {\bibfnamefont {C.}~\bibnamefont {Cocchi}},
  \bibinfo {author} {\bibfnamefont {D.}~\bibnamefont {Nabok}}, \bibinfo
  {author} {\bibfnamefont {K.}~\bibnamefont {Rezouali}}, \bibinfo {author}
  {\bibfnamefont {M.~A.}\ \bibnamefont {Belkhir}}, \ and\ \bibinfo {author}
  {\bibfnamefont {C.}~\bibnamefont {Draxl}},\ }\href {\doibase
  10.1103/PhysRevB.97.241114} {\bibfield  {journal} {\bibinfo  {journal}
  {Phys.~Rev.~B}\ }\textbf {\bibinfo {volume} {97}},\ \bibinfo {pages} {241114}
  (\bibinfo {year} {2018})}\BibitemShut {NoStop}%
\end{thebibliography}
\end{document}